\documentclass[apj]{emulateapj}
\usepackage{graphics,graphicx,times,natbib,longtable,placeins}

\def\mhalo{${\rm M_{halo}}$}
\def\mstar{${\rm M_{*}}$}
\def\msun{${\rm M_\odot}$}
\def\logm{{\rm log(M_{*}/M_\odot)}}
\def\logmh{{\rm log(M_{halo}/M_\odot)}}

\def\spose#1{\hbox to 0pt{#1\hss}}
\def\lta{\mathrel{\spose{\lower 3pt\hbox{$\mathchar"218$}}
     \raise 2.0pt\hbox{$\mathchar"13C$}}}

\shorttitle{Dwarf Metallicity}
\shortauthors{Guo et al.}

\begin{document}

\title{Stellar Mass--Gas-phase Metallicity Relation at $0.5\leq z\leq0.7$: A
Power Law with Increasing Scatter toward the Low-mass Regime}

%
\author{Yicheng Guo$^{1}$, David C. Koo$^{1}$, Yu Lu$^{2}$, John C. Forbes$^{1}$, Marc Rafelski$^{3,4}$, Jonathan R. Trump$^{5,6}$, Ricardo Amor{\'i}n$^{7}$, Guillermo Barro$^{1}$, Romeel Dav{\'e}$^{8}$, S. M. Faber$^{1}$, Nimish P. Hathi$^{9}$, Hassen Yesuf$^{1}$, Michael C. Cooper$^{10}$, Avishai Dekel$^{11}$, Puragra Guhathakurta$^{1}$, Evan N. Kirby$^{12}$, Anton M. Koekemoer$^{13}$, Pablo G. P\'erez-Gonz\'alez$^{14}$, Lihwai Lin$^{15}$, Jeffery A. Newman$^{16}$, Joel R. Primack$^{17}$, David J. Rosario$^{18}$, Christopher N. A. Willmer$^{19}$, Renbin Yan$^{20}$}
\affil{$^1$ UCO/Lick Observatory, Department of Astronomy and Astrophysics, University of California, Santa Cruz, CA, USA; {\it ycguo@ucolick.org}}
\affil{$^2$ Observatories, Carnegie Institution for Science, Pasadena, CA, USA}
\affil{$^3$ Goddard Space Flight Center, Code 665, Greenbelt, MD, USA}
\affil{$^4$ NASA Postdoctoral Program Fellow}
\affil{$^5$ Department of Astronomy and Astrophysics and Institute for Gravitation and the Cosmos, Pennsylvania State University, University Park, PA, USA}
\affil{$^6$ Hubble Fellow}
\affil{$^7$ INAF-Osservatorio Astronomico di Roma, Monte Porzio Catone, Italy}
\affil{$^8$ University of the Western Cape, Bellville, Cape Town, South Africa}
\affil{$^9$ Aix Marseille Universit{\'e}, CNRS, LAM (Laboratoire d'Astrophysique de Marseille) UMR 7326, Marseille, France}
\affil{$^{10}$ Center for Cosmology, Department of Physics and Astronomy, University of California Irvine, CA, USA}
\affil{$^{11}$ Center for Astrophysics and Planetary Science, Racah Institute of Physics, The Hebrew University, Jerusalem, Israel}
\affil{$^{12}$ California Institute of Technology, Pasadena, CA, USA}
\affil{$^{13}$ Space Telescope Science Institute, Baltimore, MD, USA}
\affil{$^{14}$ Departamento de Astrof\'{\i}sica, Facultad de CC.  F\'{\i}sicas, Universidad Complutense de Madrid, E-28040 Madrid, Spain}
\affil{$^{15}$ Institute of Astronomy \& Astrophysics, Academia Sinica, Taipei, Taiwan (R.O.C.)}
\affil{$^{16}$ Department of Physics and Astronomy, University of Pittsburgh, Pittsburgh, PA, USA}
\affil{$^{17}$ Department of Physics, University of California, Santa Cruz, CA, USA}
\affil{$^{18}$ Max-Planck-Institut f${\rm \ddot{u}}$r Extraterrestrische Physik (MPE), Garching, Germany}
\affil{$^{19}$ Steward Observatory, University of Arizona, Tucson, AZ, USA}
\affil{$^{20}$ Department of Physics and Astronomy, University of Kentucky, Lexington, KY, USA}



\begin{abstract} 

We present the stellar mass (\mstar)--gas-phase metallicity relation (MZR) and
its scatter at intermediate redshifts ($0.5\leq z\leq0.7$) for 1381 field
galaxies collected from deep spectroscopic surveys. The star formation rate
(SFR) and color at a given \mstar\ of this magnitude-limited ($R\lesssim24$ AB)
sample are representative of normal star-forming galaxies. For masses below
$10^9$\msun, our sample of 237 galaxies is $\sim$10 times larger than those in
previous studies beyond the local universe. This huge gain in sample size
enables superior constraints on the MZR and its scatter in the low-mass regime.
We find a power-law MZR at $10^{8}$\msun$<$\mstar$<10^{11}$\msun: ${\rm
12+log(O/H) = (5.83\pm0.19) + (0.30\pm0.02)log(M_{*}/M_\odot)}$. At
$10^9$\msun$<$\mstar$<10^{10.5}$\msun, our MZR shows agreement with others
measured at similar redshifts in the literature. Our power-law slope is,
however, shallower than the {\it extrapolation} of the MZRs of others to masses
below $10^9$\msun. The SFR dependence of the MZR in our sample is weaker than
that found for local galaxies (known as the Fundamental Metallicity Relation).
Compared to a variety of theoretical models, the slope of our MZR for low-mass
galaxies agrees well with predictions incorporating supernova energy-driven
winds. Being robust against currently uncertain metallicity calibrations, the
scatter of the MZR serves as a powerful diagnostic of the stochastic history of
gas accretion, gas recycling, and star formation of low-mass galaxies. Our
major result is that the scatter of our MZR increases as \mstar\ decreases. Our
result implies that either the scatter of the baryonic accretion rate
($\sigma_{\dot{M}}$) or the scatter of the \mstar--\mhalo\ relation
($\sigma_{SHMR}$) increases as \mstar\ decreases. Moreover, our measures of
scatter at $z=0.7$ appears consistent with that found for local galaxies.  This
lack of redshift evolution constrains models of galaxy evolution to have both
$\sigma_{\dot{M}}$ and $\sigma_{SHMR}$ remain unchanged from $z=0.7$ to $z=0$.
 
\end{abstract}

\section{Introduction}
\label{intro}

The relation between the gas-phase metallicity and stellar mass (\mstar) of
galaxies --- \mstar--metallicity relation (MZR) --- is one of the most
fundamental scaling relations of galaxy formation. It is a sensitive tracer of
gas inflow, consumption, and outflow, all of which regulate star formation in
galaxies. 

The MZR in the local universe has been well determined down to
\mstar$\sim10^8$\msun\ \citep{tremonti04}, and has even been explored down to
$\sim10^6$\msun\ \citep{lee06mzr}. Beyond the local universe, the MZR
measurements have been reported up to $z\sim3$
\citep[e.g.,][]{erb06mzr,maiolino08,zahid11,steidel14,maier14,sanders15,ggebhardt16}.
The MZR smoothly evolves from $z>2$ to $z=0$, with lower redshift galaxies
having higher metallicity at a given \mstar\ \citep[e.g.,][]{zahid13,pm13}.
Most of the MZR measurements beyond the local universe, however, only explore
galaxies with \mstar$>10^9$\msun\ because spectroscopically observing a
sufficiently large sample of low-mass (\mstar$<10^9$\msun) and thus faint
galaxies is very time-consuming.

The sparse number of metallicity measurements of low-mass galaxies beyond the
local universe limits our understanding of the redshift evolution of
star formation and feedback processes. First, low-mass galaxies are expected to
provide the most stringent constraints on feedback because the effect of
feedback is expected to be strong in their shallow gravitational potential
wells. Second, the scatter of the MZR of low-mass galaxies contains clues on
the stochastic nature of the formation mechanisms of low-mass galaxies
\citep{forbes14mzr}. Comparing the scatter of low-mass and massive galaxies
would tell us whether and by how much the low-mass galaxies are formed through
a more stochastic process than massive galaxies. 

Feedback processes caused by supernovae \citep{dekel86}, stellar winds
\citep{hopkins12feedback}, stellar radiation pressure \citep{murray05}, and/or
even AGN \citep{croton06} have already become essential ingredients of theories
of galaxy formation \citep[see the review of][]{somerville15}. A complete
physical picture of feedback, however, remains to be developed. In recent
years, the MZR has been widely used in analytic, semi-analytic, and numerical
models to constrain the properties of outflows
\citep[e.g.,][]{finlator08,peeples11,dave11a,dave11b,dave12,dave13,lilly13,forbes14mzr,luyu15metal}.
To understand the redshift evolution of the outflow properties, e.g., the
strength, velocity, mass loading factor, metallicity, etc., the MZR
measurements at different redshifts are needed.

The observations of the MZR of low-mass galaxies beyond the local universe are
sparse. The 26 galaxies of \citet[][H13]{henry13a} comprise the first
measurement of the intermediate-redshift MZR below ${\rm 10^9 M_\odot}$. Their
MZR is consistent with the equilibrium model with momentum-driven winds
\citep[][D12]{dave12}, but their star formation rate (SFR)--\mstar\ relation
favors, in contrast, energy-driven winds. It is possible that the equilibrium
models endure a breakdown in low-mass galaxies, but the small sample size of
H13 also urges the need of large sample to present a robust constraint on the
MZR at the low-mass end. \citet{henry13b} push the metallicity measurements of
low-mass galaxies to higher redshifts at $1.3<z<2.3$ by {\it stacking} the
Hubble Space Telescope ({\it HST})/WFC3 grism of 83 galaxies in the WISP Survey
\citep{atek10,colbert13}. Although the stacked spectrum in each \mstar\ bin has
a sufficiently high signal-to-noise ratio (S/N), it loses information of
individual galaxies.

The scatter of the MZR of low-mass galaxies beyond the local universe is also
unexplored. In the local universe, both \citet{tremonti04} and \citet{zahid12}
observed an increasing scatter as \mstar\ decreases. In contrast,
\citet{lee06mzr} found a constant scatter over a 5 dex range of \mstar, but
their sample size small. While \citet{zahid11} provides a robust measurement of
the scatter of the MZR at $z\sim0.8$ with a sufficiently large sample from
DEEP2, their sample (\mstar$>10^9$\msun) does not extend to the low-mass
regime.

In this paper, we collect data from large spectroscopic surveys in the CANDELS
fields \citep{candelsoverview,candelshst} to study the MZR and its scatter at
$0.5\leq z\leq0.7$ down to \mstar$\sim10^8$\msun. Although none of these
surveys (typically limited to $R\lesssim24.1$) is designed to study low-mass
galaxies, each still contains a sufficiently large number of low-mass galaxies.
Combining them together provides the largest sample to date in the low-mass
regime at $0.5\leq z\leq0.7$. Recently, extremely metal-poor dwarf galaxies
have been found at similar redshifts \citep[e.g.,][]{amorin14,ly15}, but the
main trend and its scatter of the MZR at \mstar$<10^9$\msun\ is still not well
determined.

\begin{table*}
\vspace*{-0.3cm}
\begin{center}
\caption[]{Data Summary
\label{tb:data}}
    \begin{tabular}{ | c | c | c | c | c | c | c | c |}
    \hline
    \hline
    Survey & Field & Instrument & Wavelength Range & Resolution  & Limiting  & Exposure & Number of Galaxies\\
           &       &            &  (\AA)           & ($R$)       &  Magnitude & time (hour) & at $0.5\leq z\leq0.7$$^a$ \\
    \hline
    \hline
    TKRS \citep{wirth04} & GOODS-N & Keck/DEIMOS & 4600--9800 & 2500 & $R\leq24.4$ & 1 & 183 (47/7) \\
    DEEP2 \citep{newman13deep2} & EGS & Keck/DEIMOS & 6500--9100 & 5000 & $R\leq24.1$ & 1 & 733 (--/--) \\
    DEEP3 \citep{cooper11deep3,cooper12deep3} & EGS & Keck/DEIMOS & 4550--9900 & 2500 & $R\leq24.1$ & 2 & 465 (128/55) \\
    \hline
    \end{tabular}
\end{center}
Note: ($a$): In each line, the number outside the bracket is the total number of galaxies in our sample. The two numbers within the bracket are the number of galaxies with $10^{8.5}$\msun$<$\mstar$\leq10^{9}$\msun and the number of galaxies with \mstar$\leq10^{8.5}$\msun, respectively. For DEEP2, we only use its 
galaxies with \mstar$>10^{9}$\msun.
\end{table*}

\begin{figure*}[htbp]
\center{\hspace*{-0.1cm}\includegraphics[scale=0.3, angle=0]{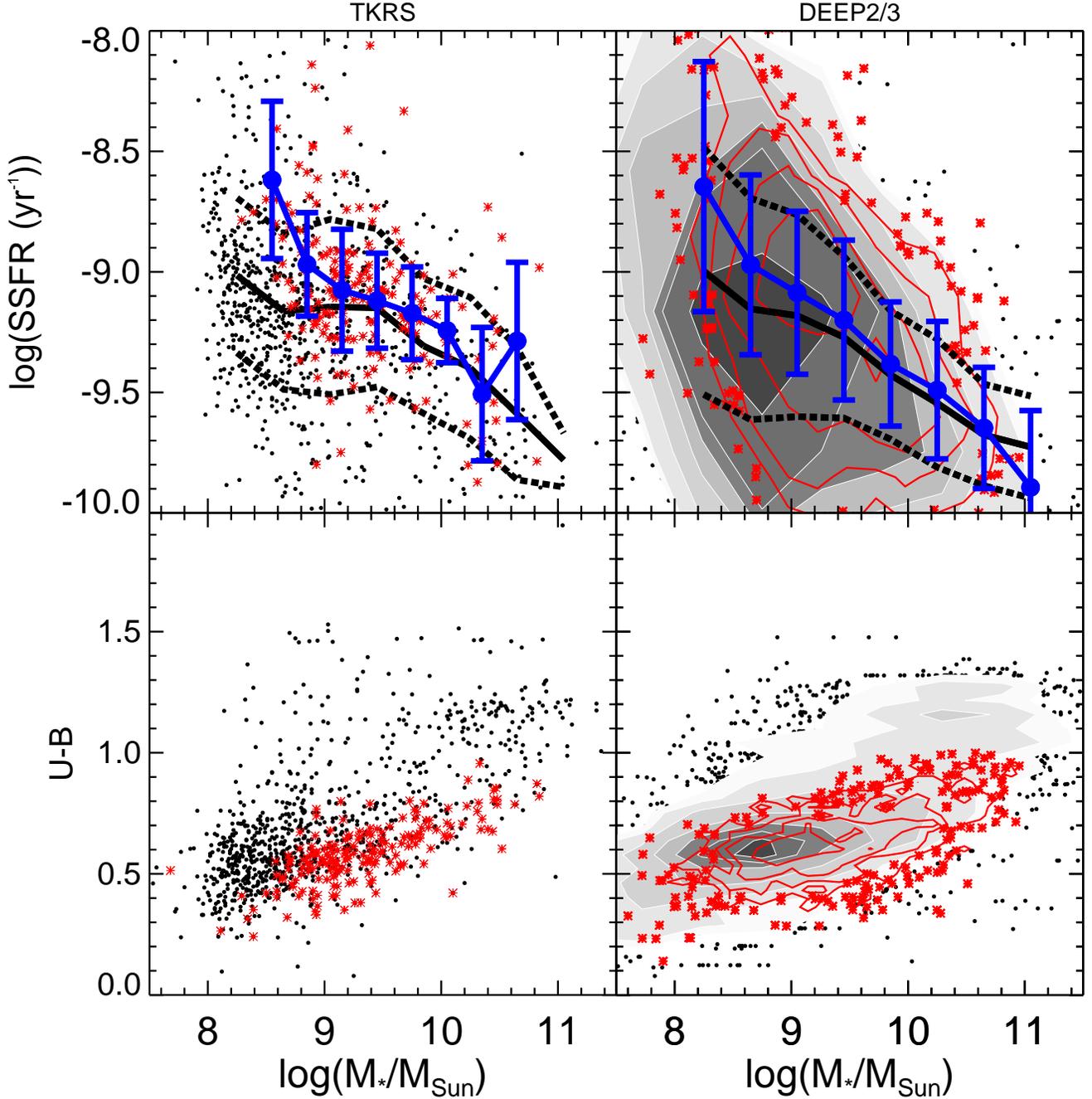}}
\vspace*{-0.2cm}

\caption[]{Sample properties. The SSFR--\mstar\ and U-B color--\mstar\
relations of the galaxies in our sample are plotted as red symbols (plus red
contours for DEEP2/3). For each survey (as shown by the title of the upper
panels), the mean and 1$\sigma$ scatter of the SSFR--\mstar\ relation of our
sample are shown by large blue circles with error bars.  Over-plotted black
dots (plus gray contours for DEEP2/3) are the mass-complete parent samples,
constructed with galaxies with $0.5\leq z\leq 0.7$ and $H<25.0$ from
CANDELS/GOODS-N (left column) catalog and $m_{3.6\mu m} < 24$ from EGS (right
column). The mean and 1$\sigma$ scatter of star-forming galaxies
(SSFR$>10^{-10}/yr$) in the parent samples, i.e., the star-forming main
sequence at $0.5\leq z\leq 0.7$, are shown by solid and dashed black lines.

\label{fig:sample}}
\end{figure*}

We adopt a flat ${\rm \Lambda CDM}$ cosmology with $\Omega_m=0.3$,
$\Omega_{\Lambda}=0.7$ and the Hubble constant $h\equiv H_0/100 {\rm
km~s^{-1}~Mpc^{-1}}=0.70$. All magnitudes are on the AB scale \citep{oke74}.
The \citet{chabrier03} initial mass function is used. The adopted solar
metallicity is 12+log(O/H)=8.69 \citep{solarz,asplund09}.

\section{Data}
\label{data}

Three deep field galaxy spectroscopic surveys are used in this paper: Team
Keck Treasury Redshift Survey \citep[TKRS;][]{wirth04}, DEEP2
\citep{newman13deep2}, and DEEP3 \citep{cooper11deep3,cooper12deep3}.
Since they are well documented and widely discussed in the literature, we only
summarize them in Table \ref{tb:data} and refer readers to the survey papers
for details. 

From each survey, our sample includes only galaxies with reliable spectroscopic
redshifts at $0.5\leq z\leq0.7$ and 3$\sigma$ [OIII] and H$\beta$ detection.
To remove AGN contamination, we exclude galaxies that fall in the upper region
(main AGN region) of the mass--excitation ([OIII]/H$\beta$ vs. \mstar) diagram
defined by \citet{juneau11}. The [OIII]/H$\beta$ metallicity indicator has
the issue of the lower--upper branch degeneracy (see Section
\ref{metal:degeneracy}). Additional emission lines are needed to break the
degeneracy. The effect of the degeneracy on the MZR, however, is believed to be
negligible for galaxies with \mstar$>$$10^9$\msun\ \citep[][and H13]{zahid11}.
Therefore, in this mass regime, we use all TKRS, DEEP2, and DEEP3 galaxies, but
at \mstar$\leq$$10^9$\msun, we only use TKRS and DEEP3 galaxies, which include
also [OII] observation to help break the degeneracy. The final sample consists
of 1381 galaxies. Among them, 273 galaxies have \mstar$<10^9M_\odot$,
comprising a sample $\sim$10
times larger than previous studies in this mass regime at similar redshift
(e.g., H13). The numbers of galaxies from each survey are listed in Table
\ref{tb:data}.

For each galaxy, we fit its multi-wavelength broad-band photometry to synthetic
stellar populating models to measure \mstar. For TKRS, we use FAST
\citep{kriek09fast} to fit the \citet{bc03} models to CANDELS multi-wavelength
catalog \citep[Barro in preparation, see also][]{ycguo13goodss}. For DEEP2/3,
we use the EGS multi-wavelength catalog of \citet{barro11a}, which is
constructed based on {\it Spitzer}/IRAC 3.6$\mu$m detection. The SED-fitting
process, detailed in \citet{barro11b}, also uses the \citet{bc03} models.

SFRs are measured by following the SFR ``ladder'' method in \citet{wuyts11a}.
This method relies on IR-based SFR estimates for galaxies detected at mid- to
far-IR wavelengths, and SED-modeled SFRs for the rest. For IR-detected galaxies
the total SFRs (SFR IR+UV) were computed from a combination of IR and
rest-frame UV luminosity (uncorrected for extinction) following
\citet{kennicutt98}. For non-IR-detected galaxies, SFRs are measured from the
extinction-corrected rest-frame UV luminosity. As shown in \citet{wuyts11a} the
agreement between the two estimates for galaxies with a moderate extinction
(faint IR fluxes) ensures the continuity between the different SFR estimates.
We refer readers to \citet{barro11b, barro13a} for the details of our SFR
measurements.

An important issue is whether our sample is representative of normal
star-forming galaxies, because it is collected from surveys which were not
designed to uniformly select low-mass galaxies. Moreover, our S/N threshold on
emission lines would introduce the Malmquist bias toward [OIII]- and
H$\beta$-bright galaxies. To test how representative our sample is, we compare
the \mstar--specific SFR (SSFR) and \mstar--color relations of our sample to
those of a mass-complete sample (parent sample) of the field of each survey at
$0.5\leq z\leq0.7$. For TKRS, we use CANDELS GOODS-N galaxies with $H<25.0$ as
the parent samples, which are approximately complete at \mstar$>10^8M_\odot$.
For DEEP2/3, we use the IRAC-detected ($m_{3.6\mu m} < 24.0$) galaxies at
$0.5\leq z\leq0.7$ from \citet{barro11b}.

Figure \ref{fig:sample} shows that our sample is actually fairly representative
of star-forming galaxies, in terms of SSFR and color at a given \mstar. The
median SSFRs of our TKRS and DEEP2/3 samples match the medians of the GOODS-N
and EGS parent samples down to \mstar$=10^9$\msun. Below $10^{9}$\msun, the
average SSFR of our sample is slightly higher than that of the star-forming
main-sequence at $0.5\leq z\leq0.7$. At \mstar$<10^9$\msun, the median SSFR of
our DEEP2/3 sample is higher than that of EGS parent sample by 0.3 dex, which
is still less than the scatter of the parent sample (0.5 dex). Since our whole
sample is dominated by DEEP2/3 sources, the comparison results of DEEP2/3
(right column of Figure \ref{fig:sample}) can be treated as being
representative of our whole sample. Overall, our sample is representative for
normal star-forming galaxies at $0.5\leq z\leq0.7$, but slightly biased toward
high-SSFR galaxies at \mstar$<10^9$\msun. This bias is also reflected in the
\mstar--color diagram, where our sample is biased toward bluer galaxies at
\mstar$<10^9$\msun. This bias mainly stems from both the R-band selection
(rest-frame $\sim$4100\AA\ at these redshifts) of each survey and our
requirement of the emission lines to be detected above the 3$\sigma$ level. In
Appendix A, we show that neither the S/N cut itself nor the bias toward
higher-SSFR galaxies introduces significant systematic offsets to our derived
MZR.

\section{Metallicity Measurement}
\label{metal}

\subsection{Line Ratio Measurement}
\label{metal:line}

\begin{figure}[htbp] 
\vspace*{-0.35cm}
\center{\hspace*{-0.35cm}\includegraphics[scale=0.32, angle=0]{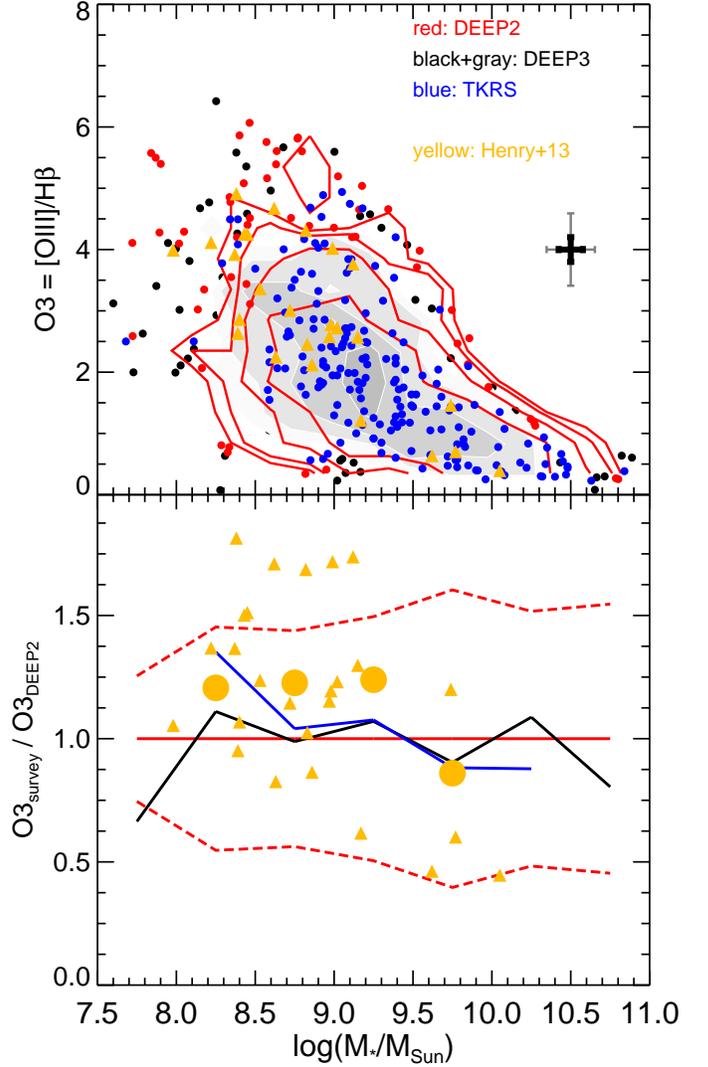}} 

\caption[]{{\it Upper}: [OIII]/H$\beta$ flux ratio as a function of \mstar\ for
galaxies in our sample. Galaxies from different surveys are shown by filled
circles with different colors. The DEEP2 (and DEEP3) distribution is shown by
red (and gray) contours plus red (and black) circles. The thick black (and thin
gray) error bars show the median (and 90 percentile) of the 1$\sigma$
uncertainty of the [OIII]/H$\beta$ and \mstar\ measurements. For comparison,
data of H13 are shown as yellow triangles. {\it Lower}: sample-to-sample
variation in the [OIII]/H$\beta$ flux ratio. We use our DEEP2 sub-sample as a
base sample and show the ratio between the average [OIII]/H$\beta$ of other
sub-samples and that of the base sample. The color scheme is the same as that
in the upper panel. The base sample (DEEP2) is shown as a constant red solid
line at unity, and its 1$\sigma$ confidence level is shown by red dashed lines.
H13 data are shown as yellow triangles, while their average values are shown as
large yellow circles.

\label{fig:raw5_raw}} 
\end{figure}

We use the [OIII]/H$\beta$ flux ratio to measure gas-phase metallicity. Unless
otherwise noted, [OIII] throughout the paper stands for [OIII]5007\AA. A
problem of using [OIII]/H$\beta$ to derive the metallicity is its dependence on
both the ionization parameter and effective temperature. Another metallicity
indicator, R23$\equiv$([OII]3727\AA+[OIII]4959,5007\AA)/H$\beta$, is often used
for optical spectroscopy at similar redshifts. Compared to [OIII]/H$\beta$, R23
has a weaker dependence on the ionization parameter, but it requires an
accurate measurement of dust extinction because the extinction is stronger for
[OII] than for [OIII]. On the other hand, [OIII]/H$\beta$ is essentially
unaffected by dust reddening because the wavelengths of [OIII] and H$\beta$ are
quite close. Because only a subset (DEEP3+TKRS) of our sample has [OII]
observed, for consistency, we only use R23 to calibrate our metallicity
measurement, but use [OIII]/H$\beta$ to derive the metallicity of the whole
sample.

To measure the fluxes of [OIII] and H$\beta$ (and [OII] if available), we
follow the steps taken by \citet{trump13}. First, a continuum is fitted across
the emission line regions by splining the 50-pixel smoothed continuum. Then, a
Gaussian function is fitted to the continuum-subtracted flux in the wavelength
regions of the emission lines. The emission line intensities are computed as
the area under the best-fit Gaussian in the line wavelength regions. To correct
for the stellar absorption of H$\beta$, we follow previous studies with DEIMOS
spectra \citep[e.g.,][and H13]{cowie08,zahid11} by assuming an equivalent width
(EW) of 1$\AA$. We then add the product of the EW and continuum to the H$\beta$
fluxes. The EW correction factor is important for our metallicity measurement.
The value of the Hb EW absorption correction depends on the spectral
resolution, and studies with lower spectral resolution typically use larger
correction factors, e.g., 3 $\AA$ used by \citet{lilly03}. In Appendix A, we
demonstrate that increasing our EW correction to 3 $\AA$ increases the
normalization of our derived MZR by 0.1 dex, but does not change its slope and
scatter.

The [OIII]/H$\beta$ flux ratio of the whole sample is shown as a function of
\mstar\ in the upper panel of Figure \ref{fig:raw5_raw}. Although galaxies from
different surveys have different spectral resolutions and exposure times, they
occupy a similar locus in the plot. A more clear view of the sample-to-sample
variation is shown in the lower panel of Figure \ref{fig:raw5_raw}. In this
panel, we use our DEEP2 sub-sample as a base sample and calculate the ratios of
the average [OIII]/H$\beta$ of individual sub-samples to that of the base
sample. The comparison results
are close to unity in most \mstar\ bins, suggesting a consistency of line-ratio
measurements between the sub-samples with different observational effects. 

Each sub-sample only shows deviation from the base sample in its lowest \mstar\
bin, where the sub-sample is subjected to small number statistics. TKRS
deviates from DEEP2 and DEEP3 at $10^{8}$\msun$<$\mstar$<10^{8.5}$\msun, where
TKRS only has 7 galaxies, while both DEEP2 and DEEP3 each have around 50
galaxies. The deviation of TKRS does not affect our conclusions in this mass
regime. At \mstar$<10^{8}$\msun, both DEEP2 and DEEP3 have only 6-7 galaxies
and hence show large discrepancy. In our analyses, we show the results of this
mass regime in plots, but do not use them in fitting or to draw any
conclusions. Overall, the consistency of line-ratio measurements suggests that
our combination of different surveys introduces no significant biases.

The galaxies of H13 (yellow triangles in both panels) provide an independent
check on our [OIII]/H$\beta$ measurements.  In general, the H13 data follow our
main trend.  Although the average [OIII]/H$\beta$ of H13 is constantly higher
than that of our DEEP2 base sample by a factor of 1.2 at
\mstar$<10^{9.5}$\msun, the deviation is still within 1$\sigma$ confidence
level of our base sample (red dashed lines). We therefore conclude that our
measurements are consistent with those in the literature.

\begin{figure}[htbp] 
\vspace*{-0.35cm}
\center{\hspace*{-0.25cm}\includegraphics[scale=0.3, angle=0]{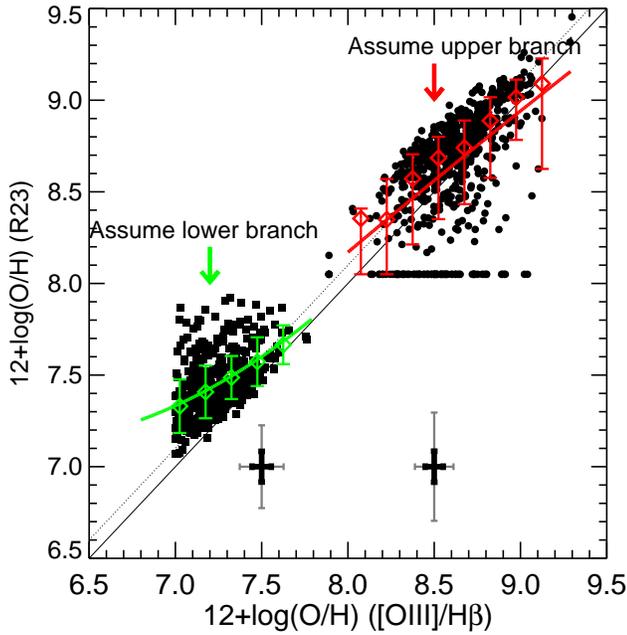}} 

\caption[]{Comparison between metallicities measured by R23 and [OIII]/H$\beta$
in the DEEP3 sample. Circles (the right locus) show the case of using the
upper-branch metallicity values for all galaxies, while squares (the left
locus) the lower-branch values. Red (green) diamonds and error bars show the
median and 16 and 84 percentiles of the comparison for the upper (lower) branch
case. The red (green) curve shows a second-order polynomial fit to the
comparison of the upper (lower) branch case. The right and left black (gray)
error bars show the median (90 percentile) measurement uncertainty of the upper
and lower branches, respectively. 

\label{fig:raw5_calib}} 
\end{figure}

\subsection{Metallicity Calibration}
\label{metal:r23}

The [OIII]/H$\beta$ flux ratio is then converted to metallicity through the
calibration of \citet[][M08]{maiolino08}. For the low-metallicity regime
(12+log(O/H)$<$8.3), M08 calibrate their relations using galaxies with
metallicities derived using the electron temperature $T_e$ method from
\citet[][]{nagao06}. For high-metallicity galaxies from SDSS DR4, they use the
photoionization models of \citet[][KD02]{kd02} to infer metallicity. A
polynomial fit is used to connect the low- and high-metallicity galaxies. 

We use R23 to calibrate our [OIII]/H$\beta$-derived metallicity. Galaxies in
DEEP3 and TKRS comprise a training sample because they have [OII] observed.
For each DEEP3 or TKRS galaxy with [OII] S/N$>$3, we measure its R23-derived
metallicity with the M08 calibration. When measuring R23, we correct for dust
extinctions of [OII], [OIII], and H$\beta$ by converting stellar continuum
extinction ${\rm E(B-V)_*}$, which is derived through the SED-fitting process
performed when measuring \mstar\ of the galaxies in our sample (see Section
\ref{data}), into gas extinction ${\rm E(B-V)_{gas}}$ through ${\rm
E(B-V)_{gas} = E(B-V)_* / 0.44}$ \citep{calzetti00}. 

The comparison between the two metallicities is shown in Figure
\ref{fig:raw5_calib}. Both [OIII]/H$\beta$ and R23 face the issue of non-unique
metallicity solutions, i.e, a given line ratio has two solutions: a
lower-metallicity value (lower branch) and a higher-metallicity one (upper
branch). For each galaxy here, we compare its solution in the same branch of
[OIII]/H$\beta$ and R23, i.e., upper vs. upper (circles in the panel) and lower
vs. lower (squares). We will break the branch degeneracy for each galaxy later.

\begin{figure}[htbp] 
\vspace*{-0.25cm}
\center{\hspace*{-0.25cm}\includegraphics[scale=0.3, angle=0]{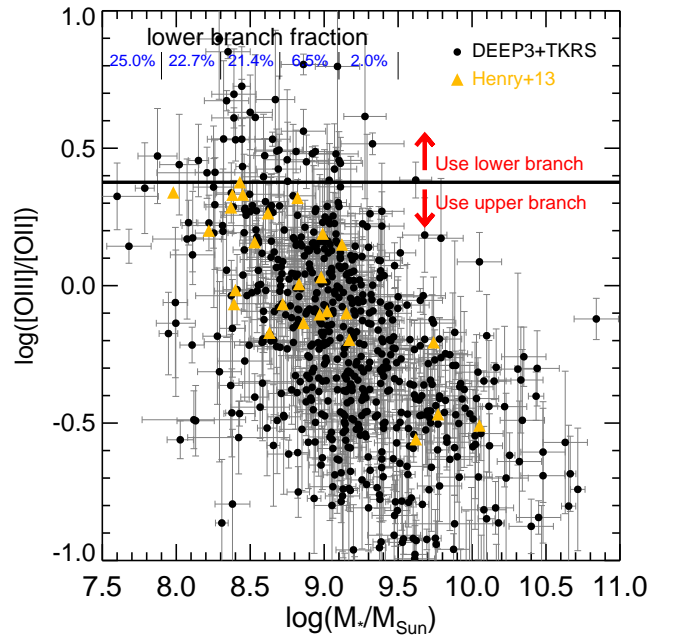}} 

\caption[]{Using [OIII]/[OII] to break the lower--upper branch degeneracy in
the DEEP3+TKRS training sample. Each black circle with gray error bars shows
one galaxy. Galaxies above log([OIII]/[OII])=0.375 (black horizontal line) are
identified to use their lower-branch metallicities, and vice versa.  In each
mass bin (width of 0.4 dex), the fraction of galaxies in the lower branch is
given by the blue number in top. For comparison, data of H13 are shown as
yellow triangles.

\label{fig:raw5_break}} 
\end{figure}

The average deviation between the two metallicities is about 0.1 dex for the
upper branch solutions. For the lower branch, the deviation increases as the
metallicity decreases, from $\sim$0.1 dex at 12+log(O/H)=7.5 to $\sim$0.3 at
12+log(O/H)=7.0. We will show later, however, that no galaxies would take a
lower-branch solution smaller than 12+log(O/H)=7.3 when we break the
degeneracy.  We fit a second-order polynomial function to the
Z(R23)--Z([OIII]/H$\beta$) relation of the upper branch (red curves) and lower
branch (green curves), respectively. We then use the best-fit relation to
correct the [OIII]/H$\beta$-derived metallicities --- both the upper- and
lower-branch solutions --- of all the galaxies in the three surveys, regardless
of whether they have [OII] observations. 

\subsection{Breaking the Upper--Lower Branch Degeneracy}
\label{metal:degeneracy}

To break the degeneracy between the two branches, we use the DEEP3+TKRS
galaxies as a training sample, where both [OII] and [OIII] are observed. As
shown by M08, [OIII]/[OII] decreases monotonically with metallicity and hence
can be used as an indicator to break the degeneracy. H13 showed that
([OIII]4959,5007\AA)/[OII]$>$3.0 works as a reasonable threshold to identify
lower-branch galaxies. The threshold of 3.0 corresponds to 12+log(O/H)$\sim$8.0
(M08), which is the turnover point of the two branches of [OIII]/H$\beta$. We
assume a flux ratio [OIII]4959\AA:[OIII]5007\AA=1:3, so our threshold is
log([OIII]/[OII])=0.375 (the horizontal line in Figure \ref{fig:raw5_break}).
Galaxies with log([OIII]/[OII])$>$0.375 are identified to be in the lower
branch, and vice versa \footnote{Here, we do not consider the measurement
uncertainties of [OIII]/[OII]. A possible caveat of ignoring the uncertainties
is that if the number of galaxies decreases dramatically as [OIII]/[OII]
increases at a give \mstar, the Eddington bias would scatter more galaxies into
the lower-branch region (above 0.375) than out of the region (below 0.375) and
hence artificially increase our lower-branch fraction. To quantify this effect
requires a large {\it complete} sample rather than a {\it representative}
sample as used in our paper.}. For each \mstar\ bin, we calculate the fraction
of galaxies that are in the lower branch. At \mstar$>10^9$\msun, only 2\% of
the galaxies are in the lower branch. This is consistent with the fact that,
although galaxies with very low metallicity have been found at $z\sim0.7$
\citep{hoyos05,amorin14,ly14,ly15}, their number densities are quite low. But
the fraction increases toward lower \mstar, and about 25\% of the galaxies at
\mstar$<10^{8.3}$\msun\ are in the lower branch. \citet{henry13b} find a
turnover of the R23--\mstar\ relation at \mstar$\sim 10^{8.5}$\msun\ in their
stacked HST/WFC3 grism spectra. Although they cannot break the degeneracy for
individual galaxies because of using the stacked spectra, their result suggests
that the lower-branch fraction becomes larger when \mstar\ decreases and that
the majority of low-mass galaxies are in the lower branch.

The above [OIII]/[OII] threshold and thus determined lower-branch fraction are
only valid for our metallicity indicator [OIII]/H$\beta$ in the M08
calibration. The ``turnover'' point of the two branches is different for
different metallicity indicators and calibrations.  For example, for
[OIII]/H$\beta$ in M08, the turnover point is around 12+log(O/H)$\sim$8.0 (see
Figure 5 of M08), but for R23 in KD02, the turnover point is around
12+log(O/H)$\sim$8.4. Therefore, galaxies with 12+log(O/H)=8.0--8.4 would be in
the upper branch of our [OIII]/H$\beta$ but in the lower branch of R23 of KD02.
This difference can explain the discrepancy between our lower-branch fraction
and that of other studies with different metallicity indicators. For example,
\citet{maier15} find a higher fraction of lower-branch galaxies at
\mstar$>10^{9.5}$\msun\ using R23 of KD02. The lowest metallicity in
\citet{maier15} is around 12+log(O/H)$\sim$8.3, which is still in the upper
branch of [OIII]/H$\beta$.  Therefore, we expect to find no object in our
[OIII]/H$\beta$ lower-branch at this mass regime. Figure \ref{fig:raw5_break}
confirms our expectation by showing no galaxies with [OIII]/[OII]$>$0.375 at
\mstar$>10^{9.5}$\msun. In this sense, our results are consistent with
\citet{maier15}.

\begin{figure*}[htbp] \vspace*{-0.8cm}
\center{\hspace*{-0.25cm}\includegraphics[scale=0.28, angle=0]{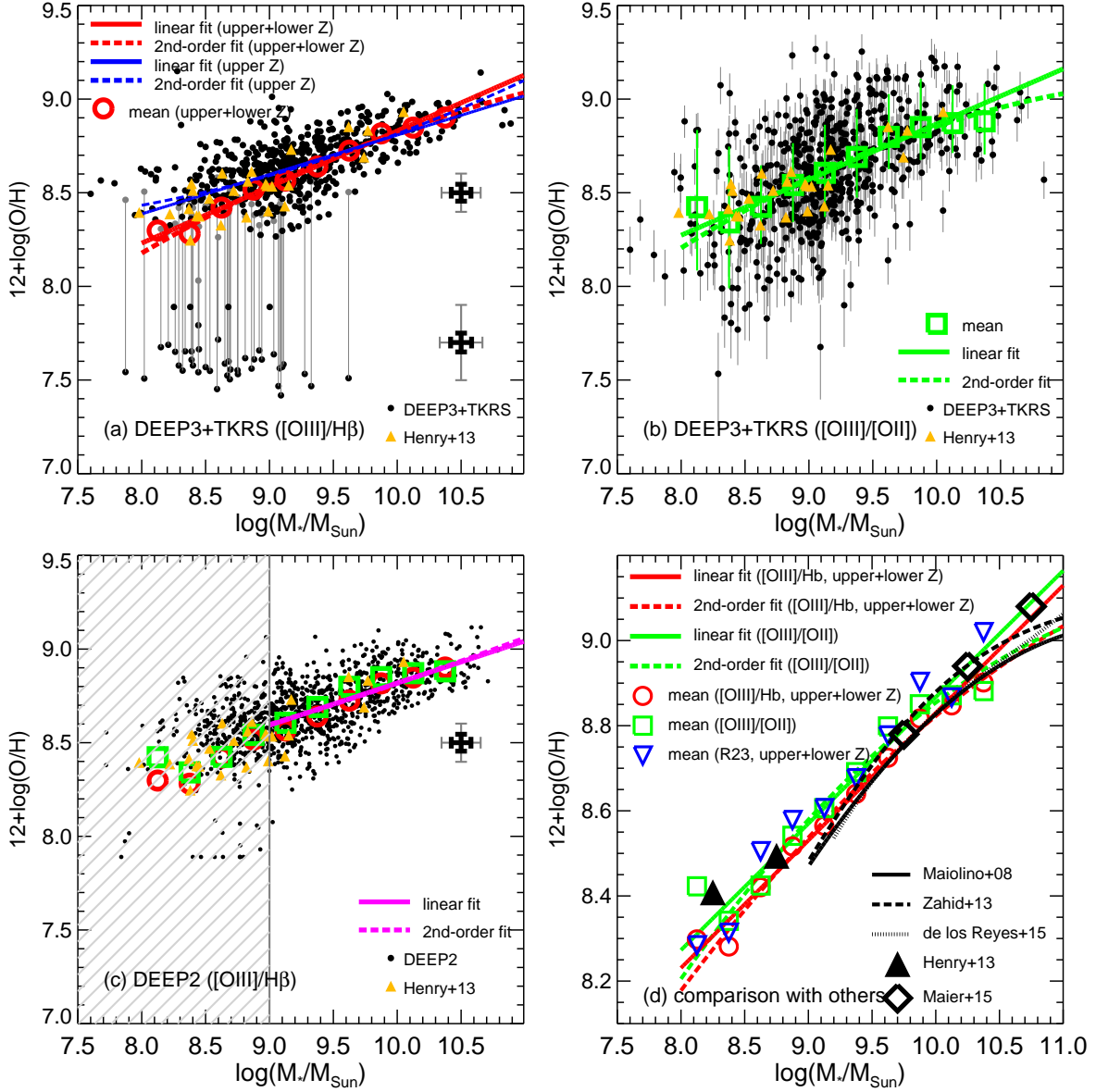}} \vspace*{-0.5cm} 

\caption[]{{\bf (a)}: MZR of our DEEP3+TKRS sample with metallicity measured
through [OIII]/H$\beta$.  Black circles are the metallicity of each galaxy
after breaking the lower--upper branch degeneracy using the threshold in Figure
\ref{fig:raw5_break}. For each lower-branch galaxy, we also show its
upper-branch metallicity by a gray circle and connect it to its lower-branch
metallicity with a gray vertical line. The means, best linear fit, and best
second-order polynomial fit of the metallicity after breaking the degeneracy
(``upper+lower Z'') are shown by red open circles, solid line, and dashed line.
As a reference, the blue solid and dashed lines show the best linear and
second-order polynomial fits of the sample using the upper-branch metallicities
(``upper Z'') for all galaxies. The upper and lower black (gray) error bars
show the median (90 percentile) measurement uncertainty of the upper and lower
branches, respectively. {\bf (b)}: MZR of our DEEP3+TKRS sample with
metallicity measured through [OIII]/[OII]. The means, best linear fit, best
second-order polynomial fit are shown by green open squares, solid line, and
dashed line. {\bf (c)}: MZR of our DEEP2 sample with metallicity measured
through [OIII]/H$\beta$.  All galaxies use their ``upper Z''. Galaxies at
\mstar$<10^{9}$\msun\ (shaded area) are significantly subjected to the branch
degeneracy and hence not used in our analyses. The best linear and
second-order polynomial fits are shown by purple solid and dashed lines. The
means of DEEP3+TKRS are shown as open circles ([OIII]/H$\beta$-derived
``upper+lower Z'') and squares ([OIII]/[OII]-derived). Data from H13, after
being converted to the M08 metallicity calibration, are shown as yellow
triangles in Panel (a), (b), and (c). {\bf (d)}: Comparison between our MZRs
and other measurements in the literature. Red and green lines and symbols are
the same as in Panel (a), (b), and (c). We also add the mean metallicity
measured through R23 in this panel (blue triangles). Black curves and symbols
show the MZRs at similar redshifts from other studies, as the labels indicate.
The y-axis range of Panel (d) is different from that of other panels.

\label{fig:mzrsub}} 
\end{figure*}

\section{Stellar Mass--Metallicity Relation at $0.5\leq z\leq0.7$}
\label{mzr}

\subsection{Stellar Mass--Metallicity Relation at $0.5\leq z\leq0.7$}
\label{mzr:mzr}

\begin{table*}
\vspace*{-0.3cm}
\begin{center}
\caption[]{Best-fit Parameters of the MZR: $12+{\rm log(O/H)} = c_0  + c_1 {\rm log(M_{*}/M_\odot)} + c_2 {\rm log(M_{*}/M_\odot)}^2$
\label{tb:param}}
    \begin{tabular}{ | c | c | c | c | c | c | c |}
    \hline
    \hline
    Sample$^a$ & Line ratio & Branch Degeneracy$^b$ & $c_0$ & $c_1$ & $c_2$  & fiducial relation$^c$ \\
    \hline
    \hline
    DEEP3+TKRS & [OIII]/H$\beta$ & upper+lower Z & $5.83\pm0.19$ & $0.30\pm0.02$ & fixed to 0 & Yes \\
    DEEP3+TKRS & [OIII]/H$\beta$ & upper+lower Z & $2.64\pm2.35$ & $0.98\pm0.51$ & $-0.04\pm0.03$ & \\
    \hline
    DEEP3+TKRS & [OIII]/H$\beta$ & upper Z & $6.70\pm0.09$ & $0.21\pm0.02$ & fixed to 0 & \\
    DEEP3+TKRS & [OIII]/H$\beta$ & upper Z & $9.48\pm1.20$ & $-0.39\pm0.26$ & $0.03\pm0.02$ & \\
    \hline
    DEEP3+TKRS & [OIII]/[OII] & --- & $5.90\pm0.18$ & $0.30\pm0.02$ & fixed to 0 & \\
    DEEP3+TKRS & [OIII]/[OII] & --- & $1.65\pm2.34$ & $1.21\pm0.51$ & $-0.05\pm0.03$ & \\
    \hline
    DEEP2 & [OIII]/H$\beta$ & upper Z & $6.57\pm0.11$ & $0.23\pm0.02$ & fixed to 0 & \\
    DEEP2 & [OIII]/H$\beta$ & upper Z & $7.53\pm2.40$ & $0.03\pm0.50$ & $0.01\pm0.03$ & \\
    \hline
    \end{tabular}
\end{center}

Note: ($a$): DEEP3+TKRS galaxies have [OII] measurements to allow breaking the
lower--upper branch degeneracy. DEEP2 galaxies have no [OII] measurements.  We
therefore only fit the DEEP2 MZR with galaxies at \mstar$\geq10^{9}$\msun,
where the fraction of galaxies in the lower branch is negligible. ($b$): This
column indicates whether we break the lower--upper branch degeneracy.
``upper+lower Z'' uses the metallicity values after breaking the lower--upper
branch degeneracy using the [OIII]/[OII] threshold in Figure
\ref{fig:raw5_break}.  ``upper Z'' does not break the degeneracy and uses the
upper-branch metallicity for all galaxies. There is no branch degeneracy issue
for the metallicity measured through [OIII]/[OII]. ($c$): The fiducial MZR
relation is used in our comparisons with other MZR measurements and models.

\end{table*}

We first measure the MZR of the DEEP3+TKRS sample, in which each galaxy breaks
the lower--upper branch degeneracy by using the [OIII]/[OII] threshold. The MZR
(called ``upper+lower Z'' hereafter) is shown in Panel (a) of Figure
\ref{fig:mzrsub}. For each galaxy with log([OIII]/[OII])$>$0.375, we connect
its lower branch (black circle) and upper branch (gray circle) metallicities
with a gray vertical line to show the difference between the two solutions. For
these galaxies, we use their lower-branch solutions when deriving the MZR,
while for other galaxies, we use their upper-branch solutions. 

We fit a polynomial function to the DEEP3+TKRS ``upper+lower Z'' MZR:
\begin{eqnarray}
\label{eq:polyfit}
12+{\rm log(O/H)} = c_0  + c_1 x + c_2 x^2,
\end{eqnarray}
where $x={\rm log(M_{*}/M_\odot)}$. This function becomes a linear function 
when $c_2=0$. The best-fit parameters, with either a free $c_2$ or a fixed
$c_2=0$, are shown in Table \ref{tb:param}.

The second-order polynomial fit (red dashed line in Panel (a)) shows almost no
difference from the linear fit (red solid line), except at the very massive
end. The $\chi^2$ value of the polynomial fit is comparable to that of the
linear fit, but an $F$-test shows that the former does not significantly
improve the goodness-of-fit by adding one more free parameter. Therefore, we
choose the linear fit as our preferred function. The linear DEEP3+TKRS MZR has
a slope of $0.30\pm0.02$. For a comparison, we also measure the MZR by using
the upper branch solutions for all galaxies. Not surprisingly, the fits (blue
solid and dashed lines and parameters of ``upper Z'' in Table \ref{tb:param})
show a flatter slope at the low-mass end.

Panel (a) shows a gap in the metallicity distribution at
\mstar$<10^{9.5}$\msun. This gap exists because we only allow each galaxy to
use one metallicity from its upper- and lower-branch values. The discreteness
of the two branches results in the gap. For example, in Panel (a), a galaxy
whose upper and lower branch metallicities are connected by a vertical gray
line can only choose either of its two endpoints but no other values. 
This gap is a generic feature of some metallicity calibrations when breaking
the degeneracy. 
It is independent on which emission lines are used to measure the metallicity.
A similar gap also exists even if we use the M08 calibration of the R23
indicator. Similar gaps from different line ratios and calibrations can be seen
from Panel (3), (4), and (5) of Figure 1 of \citet{ke08}.

Choosing one value from the two branches is a deterministic method, which
over-simplifies the metallicity calibration in the ``turnover'' region by not
taking into account of the dispersion (or uncertainty) of the calibration. For
example, the [OIII]/H$\beta$–-metallicity calibration used in our paper, namely
the ``best-fit'' relation of M08, has been shown to have a dispersion of about
0.15 dex in [OIII]/H$\beta$ at a given metallicity (Figure 5 of M08 and Figure
17 of \citet{nagao06}). This dispersion (either intrinsic or due to measurement
uncertainties), when being converted into the dispersion of metallicity at a
given emission-line ratio, would cause the emission-line ratios to lose their
diagnostic powers in the ``turnover'' region: a wide range of metallicity may
have the same emission-line ratio (see Figure 11 of \citet{ke08} for an example
of R23). In principle, this dispersion should be taken into account when
converting emission-line ratios into metallicity. This step, however, requires
that the dispersion is measured from a sample of galaxies that matches the
properties of our sample of interest. Because of the lack of such a sample at
our target redshift, we skip this step and retrograde to the deterministic
method. We then test its effect on our MZR measurement.

To test the effect of the gap, we need an indicator that changes monotonically
with metallicity to provide an independent check. [NII]/H$\alpha$ is usually
the first choice for lower-redshift galaxies, but it is shifted out of the
wavelength window of our data. Here, we use [OIII]/[OII] as the independent
check.  Although [OIII]/[OII] is in fact a diagnostic of ionization parameter,
it also provides a sort of metallicity measurement, thanks to the tight
relation between ionization parameter and metallicity \citep[see][for detailed
discussions]{nagao06}. The tight relation between ionization parameter and
metallicity is also manifested by the tight locus of star-forming main sequence
galaxies in the BPT diagrams \citep[e.g.,][]{cidfernandes07,kewley13,maier15}.
We use the calibration of M08 to convert [OIII]/[OII] to metallicity.
\citet{ggebhardt16} shows that the calibration evolves little from M08's local
sample to their z$\sim$2 sample.

The new MZR is shown in Panel (b) of Figure \ref{fig:mzrsub}. There is no gap
of metallicity in this panel, although the scatter is increased. The mean
relation and 1$\sigma$ deviation (green squares with error bars) agree very
well with the data points of H13, which are measured through R23. This
comparison ensures that using [OIII]/[OII] is able to recover the mean
metallicity at each \mstar\ bin. Also, the comparison between the best-fit
relations of the [OIII]/H$\beta$-derived and [OIII]/[OII]-derived MZRs (red and
green lines in Panel (d)) shows only small systematic offset of $\sim$0.05 dex
in normalization and no changes at all in slope. Such good agreement
demonstrates that our method of breaking the branch degeneracy introduces no
significant effects on both the slope and the normalization of the MZR. We
therefore use the ``upper+lower Z'' MZR as our fiducial one to compare with
other measurements and models later.

In Panel (d) of Figure \ref{fig:mzrsub}, we also show the mean metallicity of
each \mstar\ bin by using the R23 indicator of M08 (blue triangles). The
upper--lower degeneracy is also broken by using the same [OIII]/[OII] threshold
in Figure \ref{fig:raw5_break}. The mean R23 metallicity shows very small
deviation from our mean [OIII]/H$\beta$-derived metallicity. This result again
demonstrates that the slope and normalization of our MZR derived through
[OIII]/H$\beta$-derived are robust.

To provide a better statistics, we also measure the MZR of our DEEP2 sample
through [OIII]/H$\beta$. Since we do not have [OII] observation for DEEP2
galaxies, we are not able to break the branch degeneracy. We therefore only
measure the DEEP2 MZR down to $10^{9}$\msun\ because, as shown by the
lower-branch fraction of DEEP3+TKRS in Figure \ref{fig:raw5_break}, the number
of lower-branch galaxies is negligible in the massive and intermediate-mass
regimes. For galaxies at \mstar$>10^{9}$\msun, the DEEP2 mean MZR matches that
of DEEP3+TKRS (both [OIII]/H$\beta$ and [OIII]/[OII] derived) very well (Panel
(c) of Figure \ref{fig:mzrsub}).

\subsection{Comparison with Other MZRs}
\label{metal:compare}

Our fiducial MZR (``upper+lower Z'') at \mstar$>10^{9}$\msun\ shows good
agreement with other measurements at similar redshifts in the literature.
Panel (d) of Figure \ref{fig:mzrsub} shows the MZRs of M08, \citet{zahid13},
\citet{dlR15}, and \citet{maier15}, all converted to the calibration of KD02
(the same calibration used in M08 for high-metallicity galaxies) using the
calibration conversion table of \citet{ke08}. At
$10^9$\msun$<$\mstar$<10^{10.5}$\msun, other MZRs show a deviation of only
$<$0.05 dex from ours, which is much smaller than the scatter of our sample. At
\mstar$<10^9$\msun, the average metallicity of H13 (also converted to the
calibration of M08) is higher than our best-fit MZR by $\sim$0.1 dex at
\mstar$<10^{8.5}$\msun, which can be attributed to the fact that H13
assumes all galaxies being in the upper branch.

Although the absolute metallicity values of our sample in the intermediate
\mstar\ range match other studies, the slope of our MZR is different from that
of others. Because we prefer a linear fit, our slope is constant
($0.30\pm0.02$) across the 3-dex range of the \mstar. This result appears in
contrast to the ``saturation'' of metallicities in massive galaxies found by
other authors, e.g., M08, \citet{zahid13}, and \citet{dlR15}, who found that
the slope of MZR decreases significantly as \mstar\ increases. The different
slopes between our and other studies at the massive end could be attributed to
a few reasons.

First, we do not have enough massive galaxies at \mstar$>10^{10.5}M_\odot$ to
constrain the slope at the massive end (only 19 in our whole sample). Second,
the slope of the massive end is sensitive to AGN removal. AGN contamination
would bias the average metallicity of massive galaxies toward lower values 
because the strong [OIII] emission of AGN host galaxies would make the galaxies
resemble a low-metallicity galaxies with higher [OIII]/H$\beta$. We use the
mass-excitation method to exclude both X-ray and non-X-ray AGN, but some
studies, e.g., \citet{zahid13}, only exclude X-ray AGN (see Appendix B for
detailed discussions of AGN removal). Third, the analytic functions used to fit
the MZR are different. Other studies use second-order polynomial functions (M08
and \citet{dlR15}) or power-law \citep{zahid13} to fit the MZR in logarithmic
space. As show in Table \ref{tb:param}, the second-order polynomial fits to our
[OIII]/H$\beta$ ``upper+lower Z'' and [OIII]/[OII] MZRs have $c_2 < 0$,
indicating a slight ``saturation'' at \mstar$>10^{10.5}M_\odot$. These
polynomial fits (red and green dashed lines in Panel (d)) actually match the
MZR of M08 and \citet{dlR15} very well.

At the low-mass end, we find a high average metallicity compared to the {\it
extrapolation} of other MZR relations to \mstar$\sim10^8M_\odot$ (i.e.,
extrapolating all black lines in Panel (b) of Figure \ref{fig:mzrsub} to low
mass).
The slope and normalization of our MZR at the low-mass end do not significantly
depend on the choice of fitting functions. Other studies have almost no data at
\mstar$<10^9$\msun\ to constrain the slope at the low-mass end. The good
agreement between our MZR and that of H13 at the low-mass end (Panel (d) of
Figure \ref{fig:mzrsub}) provides reassurance to our measurement. We therefore
believe that simply extrapolating other MZRs to the low-mass regime would
underestimate the average metallicity of galaxies with \mstar$<10^9$\msun.

Our sample selection would not bias our MZR at the low-mass end. As shown in
Section \ref{data}, our sample is quite representative in terms of SSFR and U-B
color at a given \mstar\ down to \mstar$\sim10^{9}M_\odot$. Below that, our
sample is biased toward galaxies with higher SSFR (and hence bluer color). If
the local SFR dependence, i.e.,at a given \mstar, lower SFR galaxies having
higher metallicity, is also found in our sample, we would expect that our
average metallicity at \mstar$<10^{9}M_\odot$ is underestimated instead of
overestimated, which suggests that the true MZR may be even flatter than what
we find. 
As shown in Section \ref{metal:sfr}, however, we only find a very weak SFR
dependence of metallicity in our sample, which would not significantly bias our
MZR measurement. In fact, detailed tests in Appendix A show that sample
selection (in SSFR and emission-line S/N) has no significant effects on our
derived MZR. Our sample selection would bias the MZR toward higher metallicity
only if a positive SFR dependence, i.e., galaxies with higher SFR having higher
metallicity \citep{ly14}, holds at $0.5 \leq z \leq 0.7$. 

\begin{figure}[htbp] \vspace*{-0.3cm}
\center{\hspace*{-0.25cm}\vspace*{-0.1cm}\includegraphics[scale=0.30,angle=0]{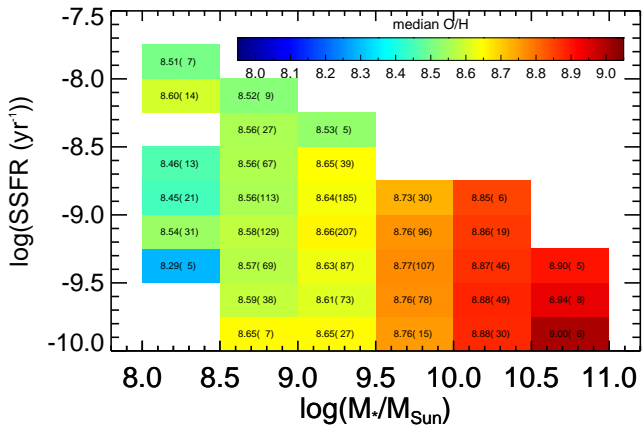}} 

\caption[]{SSFR dependence of the MZR. The median metallicity of each (\mstar,
SSFR) bin is shown by color, with the color bar in the upper right part. The
median metallicity value as well as the number of galaxies in each bin are also
given. Each bin spans 0.5 dex in \mstar\ and 0.25 dex in SSFR.

\label{fig:fmr}} 
\end{figure}

\subsection{Weak SSFR Dependence}
\label{metal:sfr}

Local galaxies are shown to follow a fundamental metallicity relation
\citep{mannucci10,mannucci11}: at a given \mstar, galaxies with higher SFR have
lower metallicity, and vice versa. The \mstar--SFR--metallicity surface (or
fundamental metallicity plane, FMR) can be collapsed to a two-dimensional space
by relating metallicity to a linear combination of \mstar\ and SFR.
\citet{mannucci11} demonstrate that metallicity correlates tightly with the
quantity $\mu_{0.32} = \log(M_{*}) - 0.32*\log(SFR)$ (in solar units) down to
\mstar$\sim 10^{9.2}$\msun. To test the SFR dependence of the MZR in our
sample, we measure the median metallicity of each (\mstar, SSFR) bin and show
the results in Figure \ref{fig:fmr}. At a given \mstar, SSFR and SFR are
equivalent. We choose the SSFR--\mstar\ space rather than the SFR--\mstar\
space because the former is used to show our sample selection effect in
Section \ref{data}. The figure is similar to and enables an easy
comparison with the representation of metallicities in the color-coded
SSFR--\mstar\ diagram in \citet{maier15} (their Figure 4). We use the
upper-branch [OIII]/H$\beta$-derived metallicities of all DEEP3+TKRS and DEEP2
galaxies for this test. Since we only present the median metallicity, given the
small fraction of lower-branch galaxies ($<$30\% in even the lowest \mstar\
bin), not identifying lower-branch galaxies would not affect the median
metallicities.

The SSFR (or SFR) dependence of metallicity in our sample is weak. Our
strongest signal is from galaxies with
$10^{8.5}$\msun$<$\mstar$<$$10^{9.5}$\msun.  In this regime, the metallicity
decreases by at most 0.15 dex when log(SSFR/yr) increases from -10 to -8. As a
comparison, \citet{mannucci10} (see their Fig. 6) find that for local galaxies
at the same \mstar\ range, the metallicity changes by 0.3 dex from
log(SSFR/yr)=-10 to -9 and would change by $\sim$0.6 dex from log(SSFR/yr)=-10
to -8 with extrapolation.

Our result is consistent with that of \citet{dlR15} and \citet{pm13}, both in
favor of a moderate or no SFR dependence of the MZR at similar redshifts.  In
contrast, some other studies \citep[e.g.,][]{cresci12,maier15} find a SFR
dependence at $z\sim0.7$ as strong as in local galaxies (i.e., a non-evolved
FMR between $z\sim0$ and $z\sim0.7$). The discrepancy indicates the uncertainty
of the existence of a fundamental metallicity relation beyond the local
universe. The weaker SFR dependence could be a physical phenomenon at
intermediate redshift, suggesting that galaxies need time to establish the SFR
dependence of their metallicity. This speculation is consistent with the
results at even higher redshifts from \citet{steidel14}, \citet{sanders15}, and
\citet{ggebhardt16}, who find no SFR dependence of the MZR at z$\sim$2.  But
the lack of SFR dependence could also be due to selection effects. Mock
observations by using local galaxies with strong SFR dependence to mimic
high-redshift observations are needed to test the selection effects
\citep{salim15}.

\begin{figure*}[htbp]
\vspace*{-0.30cm}
\center{\hspace*{-0.5cm}\includegraphics[scale=0.32, angle=0]{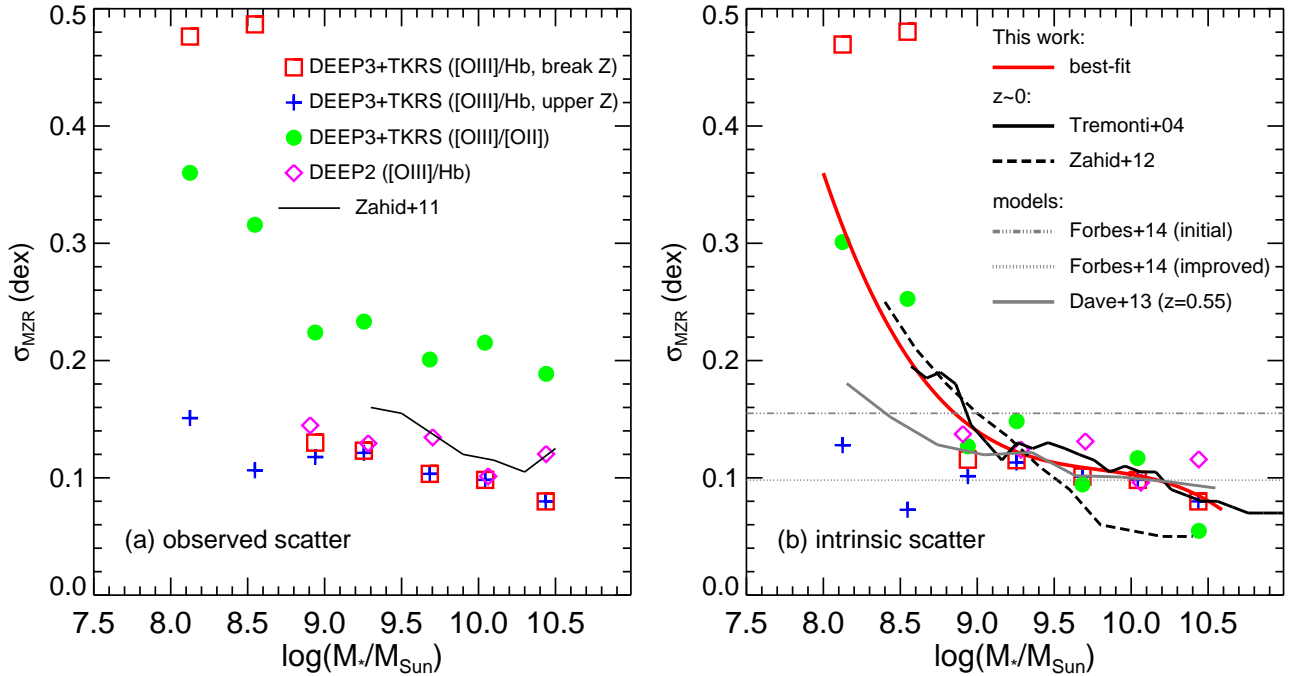}}
\vspace*{-0.5cm}

\caption[]{Scatter of the MZR of our sample. Scatter is defined as the
half-width between the 84th and 16th percentiles (${\rm P_{84}}$ and ${\rm
P_{16}}$) of the metallicity distribution at a given \mstar. {\bf Left (a)}:
Observed scatter. Measurements from different samples and/or different methods
of breaking the lower--upper branch degeneracy are plotted with different
symbols as the labels show. We only measure the scatter of DEEP2 down to
\mstar=$10^{9}$\msun. The solid black line shows the scatter of \citet{zahid11}
at $z\sim0.8$ from DEEP2 data. {\bf Right (b)}: Intrinsic scatter. The
measurement uncertainties are subtracted from the observed scatter in
quadrature. Symbols have the same meaning as in Panel (a). A third-order
polynomial function is fitted to our measurements: red squares, green circles,
and purple diamonds at \mstar$\geq10^{9}$\msun\ and only green circles at
\mstar$\leq10^{9}$\msun. The scatter of local MZRs of \citet{tremonti04} and
\citet{zahid12} are shown by solid and dashed black lines. The scatter from the
initial and improved guesses of \citet{forbes14mzr} are shown by dashed--dotted
and dotted gray lines. The scatter of numerical simulations of \citet{dave13}
is shown by the solid gray line.

\label{fig:scatter}}
\end{figure*}

\section{Scatter of the MZR}
\label{scatter}

Our $\sim$10-fold gain in sample size enables a solid study of the scatter of
the MZR at the low-mass end at $0.5 \leq z \leq 0.7$ for the first time. Figure
\ref{fig:scatter} shows the scatter of the MZR of our sample. We define the
scatter as the half width between the 84th and 16th percentiles (${\rm P_{84}}$
and ${\rm P_{16}}$) of the metallicity distribution at a given \mstar. 

We measure the scatter for our four MZRs: (1) DEEP3+TKRS
[OIII]/H$\beta$-derived ``upper+lower Z'', (2) DEEP3+TKRS
[OIII]/H$\beta$-derived ``upper Z'', (3) DEEP3+TKRS [OIII]/[OII]-derived, and
(4) DEEP2 [OIII]/H$\beta$-derived. The results are shown by different symbols
in Figure \ref{fig:scatter}. The DEEP3+TKRS [OIII]/H$\beta$-derived ``upper Z''
scatter can be treated as a lower limit of the MZR scatter because putting all
galaxies in the upper branch artificially reduces the MZR scatter. The
DEEP3+TKRS [OIII]/H$\beta$-derived ``upper+lower Z'' scatter shows a
significant upward jump at \mstar$<10^{9}$\msun, which is an artificial result
of the metallicity gap seen in Panel (a) of Figure \ref{fig:mzrsub}. We only
measure the DEEP2 scatter at \mstar$\gtrsim10^{9}$\msun, where we believe the
effect of lower branch is negligible. The DEEP2 scatter agrees with the two
[OIII]/H$\beta$-derived DEEP3+TKRS scatter. The [OIII]/[OII]-derived DEEP3+TKRS
scatter is about 0.1--0.15 dex higher than others. This large discrepancy is
due to the larger uncertainty of our [OII] measurement than [OIII] and
H$\beta$. In addition to the spectra S/N, [OII] also includes the uncertainty
(both random and systematic) of dust extinction E(B-V) measurements. The
typical uncertainty of E(B-V) in our SED-fitting is about 0.1--0.15 mag.

Our [OIII]/H$\beta$-derived scatter is slightly smaller than that of
\citet{zahid11} at $10^{9.5}$\msun$<$\mstar$<10^{10.5}$\msun\ by $\sim$0.03
dex.  We do not correct for the measurement uncertainty for both studies.
\citet{zahid11} measure the MZR of DEEP2 galaxies at $z\sim0.8$. Since the data
of their and our studies are quite similar in terms of instrument, resolution,
and exposure time, the agreement (albeit the small difference) reassures us the
accuracy of our scatter measurement. But our sample extends the scatter
measurement down to \mstar$\sim10^8$\msun, 10 times below the \mstar\ limit
adopted by \citet{zahid11}.

The intrinsic scatter of the four MZRs is shown in Panel (b), with the same
symbols as in Panel (a). To obtain the intrinsic scatter, we subtract the
average measurement uncertainty in quadrature from the observed scatter at each
\mstar\ bin. To unify the scatter from the four MZRs, we fit a third-order
polynomial function to them. We do not include the [OIII]/H$\beta$-derived
DEEP3+TKRS ``upper Z'' scatter because it serves as the lower limit. We also
do not include the [OIII]/H$\beta$-derived DEEP3+TKRS ``upper+lower Z'' scatter
at \mstar$<10^{9}$\msun because it is affected by the metallicity gap of
breaking the degeneracy. These exclusions leave the [OIII]/[OII]-derived
DEEP3+TKRS scatter as the only constraint at \mstar$<10^{9}$\msun.

The intrinsic scatter increases as \mstar\ decreases. The scatter starts as
$<$0.1 dex at \mstar$>10^{10}$\msun, gradually increases to 0.15 dex at
\mstar$\sim10^9$\msun, and then quickly increases to 0.3 dex at
\mstar$\sim10^8$\msun. The dramatic increase at the low-mass end is boosted by
the long tail of very low metallicity galaxies (Panel (b) of Figure
\ref{fig:mzrsub}). This low-metallicity tail also indicates that the
metallicity distribution at the low-mass end is skewed toward low-metallicity
galaxies \citep[see][for similar result of local galaxies]{zahid12}.

We compare the scatter of our MZR (the best fit one, i.e., the red solid line
in Panel(b)) to that of other studies (also measured as $({\rm P_{84}}$-${\rm
P_{16}})/2$). When comparing with the scatter of local MZRs, we assume that the
metallicity measurement uncertainty in the studies of local galaxies is
negligible compared to the MZR scatter \citep[e.g.,][]{zahid12}. Therefore, we
do not correct for the measurement uncertainty for their scatter.
\citet{tremonti04} measure the scatter of the local MZR of $\sim$50,000
galaxies. Their scatter (solid black line in Panel (b)) matches ours
excellently from \mstar=$10^{8}$\msun\ to \mstar=$10^{10.5}$\msun.
\citet{zahid12} re-visit the scatter of local galaxies by using $\sim$20,000
SDSS galaxies plus $\sim$800 DEEP2 galaxies to explore the faint luminosity
regime. Their scatter matches that of \citet{tremonti04} and ours very well at
\mstar$<10^{9.5}$\msun, but their scatter decreases faster than ours when
\mstar\ increases. One possible reason of their smaller scatter is that, as
\citet{zahid12} argued, their [NII]/H$\alpha$-derived metallicity is saturated
at high metallicities.

Overall, we present the first measurement of the scatter of the MZR down to
\mstar$\sim10^{8}$\msun\ at $0.5 \leq z \leq 0.7$. The scatter increases as
\mstar\ decreases, due to an increase in a low-metallicity tail of galaxies.
The scatter of the MZR shows no evolution from $z\sim0.7$ to $z\sim0$,
especially for low-mass galaxies. In Section \ref{model:scatter}, we will
discuss how to
use the scatter to shed light on the formation mechanisms of low-mass galaxies.

\section{Comparing Models to Observations}
\label{model}

\subsection{Calibration Uncertainties}
\label{model:calib}

The uncertainty of the metallicity calibration needs to be taken into account
when we compare models to observations. In this paper, to map the measured
emission line ratios to metallicities, we use the M08 calibration. There are,
however, about a dozen calibrations available in the literature. \citet{ke08}
show that, for local galaxies, the MZRs derived by using different calibrations
show significant discrepancy, with the normalization at the massive end varying
by 0.7 dex. Currently, it is not clear which of the calibrations is the most
accurate one, therefore, using any single calibration to derive the MZR for
comparison with models does not include the uncertainty in the calibration of
mapping line ratios to metallicities. 

To explore the effect of the calibration uncertainty, we convert our fiducial
MZR from the M08 calibration to the other seven calibrations discussed in
\citet{ke08}. Although \citet{ke08} does not provide the conversion from M08 to
others, since M08 uses KD02 for its upper branch calibration, we use the
conversion between KD02 and others in \citet{ke08} for this purpose. The
converted MZRs show a significantly large discrepancy (gray shaded areas in
Figure \ref{fig:theo3p}). The discrepancy of the normalization is small
($\sim$0.1 dex) at the low-mass end, but large ($\sim$0.4 dex) at the massive
end. We also show the calibration uncertainty for the slope of the converted
MZRs as the gray area in Panel (a) of Figure \ref{fig:theo3p}. The smallest
discrepancy of the slope 
occurs around \mstar$\sim10^{10}$\msun, while the discrepancy at the low-mass
end is large.

\subsection{Comparisons with Different Models}
\label{model:model}

\subsubsection{Simple Scaling Relation: \citet{dekel03}}
\label{model:dekel}

\citet{dekel03} present a simple model to study the role of feedback in
establishing basic scaling relations of low-surface brightness and dwarf
galaxies. In an instantaneous-recycling approximation, the model assumes that
the amount of metal produced in a gas-rich galaxy is proportional to the
fraction of gas that makes stars in the disk: $Z \propto \eta_{gas} \equiv
M_{*} / M_{gas}$. The model also assumes that the supernova energy required to
heat the gas is proportional to the final \mstar: $E_{SN} =
\frac{1}{2}M_{gas}V^2 \propto M_{*}$, where $V$ is the Virial velocity. $V$ is
related to the dynamical mass of the system via: $V \propto M_{dyn}^{1/3}$.
Combining all above relations, one gets: 
\begin{equation}
\label{eq:dekelz}
Z \propto \eta_{gas} \propto M_{*} / M_{gas} \propto V^2
\propto M_{dyn}^{2/3}. 
\end{equation}
In a gas-rich system, $M_{dyn} \sim M_{gas}$, so $M_{*} / M_{gas} \propto M_{dyn}^{2/3}$ yields $M_{dyn} \propto M_{*}^{3/5}$, in which case
\begin{equation}
\label{eq:dekelslope}
Z \propto M_{*}^{2/5}
\end{equation}
 --- an MZR with a slope of 0.4 in logarithmic space.

The MZR slope of this idealized model (light brown line in Panel (a) of Figure
\ref{fig:theo3p}) is slightly steeper than the slope of our best-fit linear MZR
(red line). The agreement indicates that supernova feedback could play a
primary role in determining the MZR for low-surface brightness and dwarf
galaxies. This model is, however, not valid for high luminosity and
high-surface brightness galaxies (e.g., \mstar$>10^{10}$\msun), for which
\citet{dekel03} argues $Z \sim $ constant.

\subsubsection{Slope of the MZR in Equilibrium Models: \citet{dave12}}
\label{model:winds}

Recently, equilibrium models \citep[e.g.,][and D12]{finlator08,dave11a,dave11b}
provide a simple and effective way to understand the connections between galaxy
scaling relations and physical parameters. These models are constructed based
on an assumption of the equilibrium between gas inflow, consumption, and
outflow. In these models, metallicity can be written as (D12):
\begin{equation}
\label{eq:z}
Z = \frac{y}{1+\eta}\frac{1}{1-\alpha_Z},
\end{equation}
where $y$ is the oxygen yield, $\eta$ is the mass loading factor of outflows,
and $\alpha_Z=Z_{in}/Z_{ISM}$ is the ratio of the metallicities of infalling
gas and the ISM. Oxygen abundance is then
\begin{eqnarray}
\label{eq:zoh}
12 + \log(O/H) = 12 + \log(Z) - \log(\frac{3}{4} \frac{M_O}{M_H}) \\
     = \log(y) - \log(1+\eta) + \log(\frac{1}{1-\alpha_Z}) + C, 
\end{eqnarray}
where ${M_O}$ and ${M_H}$ are the atomic mass of oxygen and hydrogen,
respectively, and $C$ is a constant.

To derive the absolute value of metallicity, the value of yield $y$ is
necessary \footnote{There are two types of the definition of $y$. First,
\citet{searle72} defined $y$ as the rate of metals produced and ejected divided
by the NET rate at which H is removed from the ISM. Then, \citet{tinsley78}
defined $y$ as metals ejected per unit mass of new stars formed. Basically, the
former defines $y$ as metal mass returned per unit long lived stars formed
(stellar mass), while the latter defined $y$ as metal mass returned per unit
star formation. In the equilibrium models discussed in our paper, the second
definition is used.}. The value depends on the age, metallicity, and IMF of the
stellar population as well as nucleosynthetic models. In current models, the
oxygen yield is between $0.008 < y < 0.021$ \citep{finlator08}. This range
translates into an uncertainty of 0.3 dex in the predicted metallicity, which
makes the comparison of the absolute MZRs difficult. Therefore, we only compare
the slope of the MZRs here. Some authors use $y$ as a free parameter to
normalize the MZR. This should be done with caution because once the age,
metallicity, and IMF of stellar population is fixed, the yield is also fixed,
and varying the yield means leaking (or producing) a fraction of metal to (or
from) nowhere.

The slope of the MZR in equilibrium models can be re-expressed in terms of the halo mass (\mhalo) as
\begin{equation}
\label{eq:slope}
\beta_{MZR} = \frac{\partial (OH)}{\partial x} = \frac{\partial (OH)}{\partial x_h} \times \frac{\partial x_h}{\partial x},
\end{equation}
where $(OH)=12+\log(O/H)$, $x=\logm$, and $x_h=\logmh$.
The term ${\partial x_h}/{\partial x} \equiv \Upsilon$ is determined by the
\mhalo--\mstar\ relation \citep[e.g.,][]{moster10,behroozi13} and absorbs all dependence
of Z on \mstar. We can now parameterize both $\eta$ and $\alpha_Z$ as a function
of \mhalo\ only. Following D12, we have
\begin{equation}
\label{eq:eta}
\eta = (10^{x_h}/10^{12})^{-\gamma},
\end{equation}
where $\gamma=1/3$ for the momentum-driven wind and $\gamma=2/3$ for the
energy-driven wind. For the term with $\alpha_Z$, we assume\footnote{In D12,
$\alpha_Z$ is expressed as a function of \mstar. Here we re-define it as a
function of \mhalo\ because all \mstar\ dependence is absorbed by $\Upsilon$.
Also, there is an error in the $\alpha_Z$ definition in D12. The
correct formula should be $\log(1/(1-\alpha_Z)) = (0.5-0.1z)(M_* / 10^{10}
M_\odot)^{0.25}$ (R. Dav{\'e}, private communication).  Equation (9) of H13 used
the incorrect formula by following D12.}, we assume
\begin{equation}
\label{eq:az}
\log(\frac{1}{1-\alpha_Z}) = (0.5-0.1z)(10^{x_h}/10^{12}) ~{\rm at}~ x_h<12
\end{equation}
and
\begin{equation}
\label{eq:az2}
\log(\frac{1}{1-\alpha_Z}) = (0.5-0.1z) ~{\rm at}~ x_h \ge 12.
\end{equation}
It is important to note that this parameterization of the
$\alpha_Z$ term is a crude approximation of the result of \citet{dave11b}.

Combining above equations, we have an expression of 
$\beta_{MZR}$:
\begin{equation}
\label{eq:beta_low}
\beta_{MZR} = [\gamma \frac{\eta}{1+\eta} + \ln(10)(0.5-0.1z)10^{(x_h-12)}] \Upsilon
\end{equation}
for $x_h<12$ and 
\begin{equation}
\label{eq:beta_high}
\beta_{MZR} = [\gamma \frac{\eta}{1+\eta}] \Upsilon
\end{equation}
for $x_h > 12$. 

When $x_h \ll 12$, $\eta \gg 1$ and $10^{(x_h-12)} \ll 1$, we have $\beta_{MZR}
\sim \gamma \Upsilon$. In this mass regime, $\Upsilon \sim 0.5$ (see the
\mhalo--\mstar\ relation of \citet{moster10}), so $\beta_{MZR} \sim 0.5\gamma$.
For a momentum-driven wind, $\beta_{MZR} \sim 0.17$, and for an energy-driven
wind, $\beta_{MZR} \sim 0.33$. The latter shows an excellent agreement with our
best-fit linear MZR ($0.30\pm0.02$).

When $x_h \gg 12$, $\eta \sim 0$, so $\beta_{MZR} \sim 0$, implying a constant
$Z$ for very massive galaxies, consistent with the argument of \citet{dekel03}.

Between the two extreme cases (e.g., $11<x_h<12$), both the $\eta$ and the
$\alpha_Z$ terms contribute to the slope ($\beta_{MZR}$). Moreover, as $x_h$
increases, the contribution of the $\alpha_Z$ term becomes larger. Therefore,
the slope of models depends significantly on the assumed $\alpha_Z$ term.
Given our crude approximation here, our comparison in this halo mass (or its
corresponding \mstar) range is very uncertain.

\begin{figure}[htbp]
\vspace*{-0.20cm}
\center{\hspace*{-0.5cm}\includegraphics[scale=0.27, angle=0]{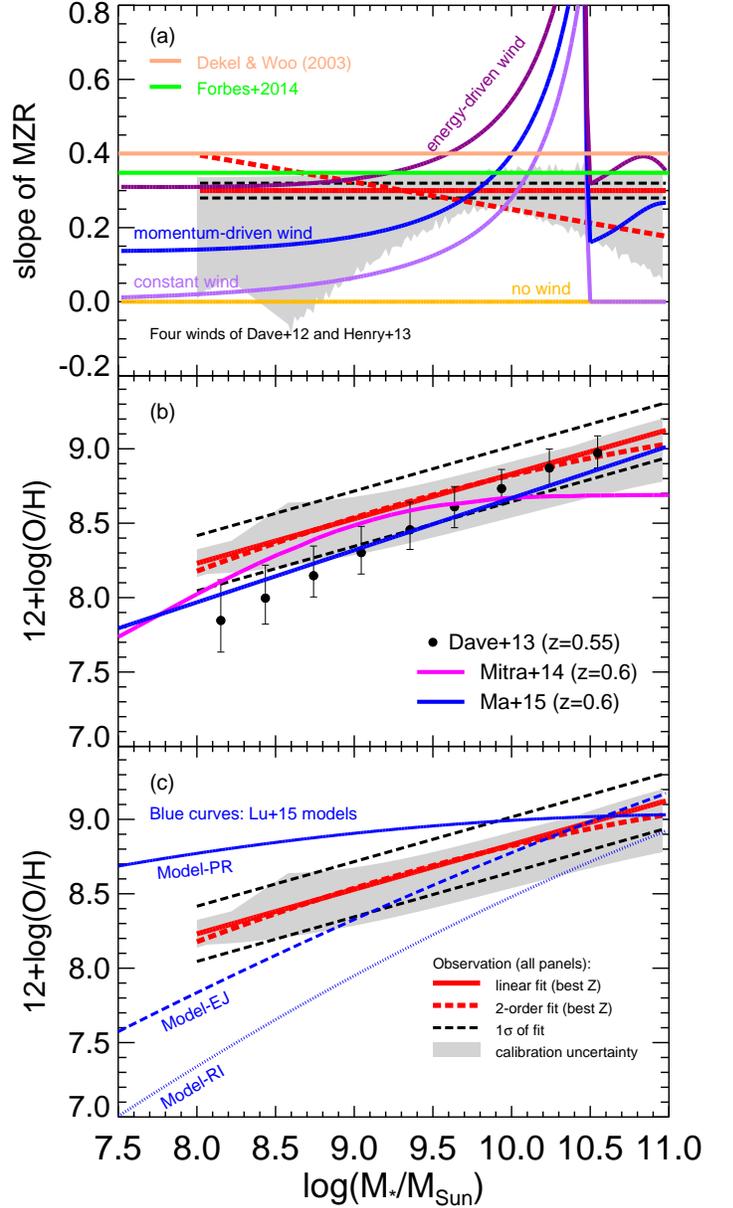}}

\caption[]{Comparisons between our MZR and others from theoretical works. In
all panel, red solid and dashed lines are best-fit linear and second-order
polynomial functions of the DEEP3+TKRS sample, galaxies in which break their
lower--upper branch degeneracy by using [OIII]/[OII] (``upper+lower Z''). The
black dashed lines are the 1$\sigma$ range of the best linear fit. The gray
shaded area is the calibration uncertainties of different metallicity
calibrations (Sec. \ref{model:calib}).  {\bf Top (a)}: Slope of the MZRs.
Slopes from theoretical works, as the text shows, are over-plotted against our
slopes. The sharp discontinuity of wind models around \mstar$=10^{10.5}$\msun\
is due to the fact that $\alpha_Z$ term in Equation \ref{eq:z} becomes abruptly
independent on \mhalo\ at \mhalo$>10^{12}$\msun. {\bf Middle (b)}: MZRs from
\citet{dave13,mitra15,ma16mzr} are compared with our MZRs. The symbols with
error bars show the 16th, 50th, and 84th percentiles of the data of
\citet{dave13}. {\bf Bottom (c)}: Model MZRs of \citet{luyu15metal} are
compared with our MZRs. Blue solid, dashed, and dotted lines are for models
with preventative feedback (PR), ejective feedback (EJ), and ejective feedback
plus gas reincorporation (RI), respectively.

\label{fig:theo3p}}
\vspace*{-0.3cm}
\end{figure}

In Panel (a) of Figure \ref{fig:theo3p}, we show the slopes of different wind
models. The slopes are now calculated numerically with the \mstar--\mhalo\
relation taken from \citet{moster10}. At \mstar$<10^{9.5}$\msun, the slope of
the energy-driven wind model matches our observation very well. The
momentum-driven wind model predicts a much flatter slope than our best-fit MZR,
although it is still within the metallicity calibration uncertainty (gray
area). At $10^{9.5}$\msun$<$\mstar$<10^{10.5}$\msun, the slopes of all models
become abnormally steep. As discussed above, this is due to the crude
approximation of the $\alpha_Z$ term.  We cannot draw conclusions on the model
comparison in this mass regime, unless a more realistic parameterization of
$\alpha_Z$ is available. At \mstar$>10^{10.5}$\msun, the slope is dominated by
the $\alpha_Z$ term and sharply drops to a value similar to the slope at the
low-mass end.

\subsubsection{Equilibrium Model with Monte Carlo Markov Chain (MCMC): \citet{mitra15}}
\label{model:mcmc}

\citet{mitra15} investigate how well a simple equilibrium model can match
observations of key galaxy scaling relations from $z=2$ to $z=0$.  The
metallicity formula in their paper is similar to our Equation \ref{eq:z}:
\begin{equation}
\label{eq:zmcmc}
Z = \frac{y}{1+\eta}(1+\frac{\dot{M}_{recyc}}{\zeta \dot{M}_{grav}}),
\end{equation}
where $\dot{M}_{recyc}$ is the rate of recycling ejected gases,
$\dot{M}_{grav}$ the gas accretion rate \citep{dekel09gas}, and $\zeta$ the
preventative feedback parameters. The key baryon cycling parameters are $\eta$,
$\zeta$, and $t_{rec}$ (gas recycling timescale). They determine the free
parameters by fitting the model to observed scaling relations: \mhalo--\mstar\,
SFR--\mstar, and MZR via MCMC. Because MZR is the
scaling relation of interest in our paper, we use their best-fit model that
only uses \mhalo--\mstar\ and SFR--\mstar\ to constrain free parameters. Each
parameter is a function of \mhalo\ and redshift. Their best-fit MZR at $z=0.6$
is shown in Panel (b) of Figure \ref{fig:theo3p}, with a yield $y=0.0126$
taken from \citet{asplund09}.

In the \mstar\ range of $10^{8.5}$\msun$<$\mstar$< 10^{9.5}$\msun, their
best-fit MZR matches the normalization of our MZR. But their model
underestimates metallicity both very low-mass and massive ends.

At \mstar$<10^{8.5}$\msun, the slope of their MZR is about $\beta_{MZR} \sim
0.5$, steeper than our slope (0.30$\pm$0.02). This is because their best-fit
$\gamma$ in our Equation \ref{eq:eta} (called $\eta_3$ in their paper) is 1.16,
much larger than both the momentum-driven and energy-driven wind. According to
our discussion above, at $x_h \ll 12$, $\beta_{MZR} \sim \gamma \Upsilon$.
Given $\Upsilon \sim 0.5$ and $\gamma=1.16$, their $\beta_{MZR} \sim 0.6$. The
large $\gamma$ means the mass loading factor at the low-mass end is so large
that lots of metals are ejected out of halos, resulting in low gas-phase
metallicities.

At \mstar$>10^{9.5}$\msun, their MZR becomes flat ($\beta_{MZR} \sim 0$).
Although the ``saturation'' of metallicity has been reported in the literature
\citep[e.g.,][]{zahid13,zahid14}, it usually happens at a much higher \mstar\
(\mstar$>10^{10.5}$\msun). As we discuss above, at \mstar$>10^{9.5}$\msun,
metallicity and $\beta_{MZR}$ become more dominated by the gas recycling term.
A key parameter in this term is $\zeta$, the preventative feedback parameter.
In D12 and \citet{mitra15}, $\zeta$ is contributed by four sources:
photo-ionization, winds, gravitational heating due to structure formation, and
quenching of star formation. At $10^{9.5}$\msun$<$\mstar$< 10^{10.5}$\msun, the
gravitational heating is the dominant contribution. \citet{mitra15} uses the
formula from the hydrodynamic simulations of \citet{fg11}, which do not include
metal-line cooling. If the metal-line cooling is considered, $\zeta$ would
decrease because the heating efficiency becomes lower, which would result in a
higher metallicity. We suspect that this could be one reason why the model of
\citet{mitra15} has lower metallicity than our results at
\mstar$>10^{9.5}$\msun.

\subsubsection{Hydrodynamic Simulations with Hybrid Winds: \citet{dave13}}
\label{model:ezw}

Motivated by analytic models \citep{murray05,murray10} and hydrodynamic
simulations \citep{hopkins12feedback} of outflows from interstellar medium,
\citet{dave13} use a hybrid wind model in their cosmological hydrodynamic
simulations: in dwarf galaxies, the energy from SNe plays a dominant role in
driving outflows, while in larger systems, the momentum flux from young stars
and/or SNe is the dominant driver. As a result, the outflow scalings switch
from momentum-driven at high masses to energy-driven at low masses. The
transition occurs at galaxy velocity dispersion $\sigma = 75 km s^{-1}$
(roughly \mstar$\sim10^{9.5}$\msun). We show the 16th, 50th, and 84th
percentiles of the metallicities of different runs of their simulations in
Panel (b) of Figure \ref{fig:theo3p}.

Interestingly, the MZR of \citet{dave13} matches our MZR very well at
\mstar$>10^{9.5}$\msun, but it gradually deviates from ours below
\mstar$=10^{9.5}$\msun. This seems inconsistent with our previous discussion,
where we show that the energy-driven wind in equilibrium models matches the
low-mass end slope of our MZR very well.  Similar deviation has been found when
\citet{dave13} compare their MZR to that at $z=0$ from SDSS \citep{tremonti04},
i.e., good agreement at the massive end, but with a steeper-than-observed slope
at the low-mass end. A possible reason for the deviation from the energy-driven
wind in equilibrium model is that low-mass galaxies in the simulations are not
in equilibrium, i.e., $dM_{gas}/dt \neq 0$. Low-mass galaxies grow quite
rapidly. Even with the energy-driven wind, the mass-loading factor is not large
enough to expel enough gas mass to maintain an “equilibrium”. At decreasing
halo (or stellar) mass, the gas reservoir becomes increasingly large relative
to what it would be in equilibrium. This makes the slope of the MZR steeper
than what is expected from the equilibrium model. Another reason could be the
accuracy of the ISM physics adopted in the simulations. Limited by the
numerical resolution, most of the cosmological hydrodynamic simulations adopt
an approximate (or ``sub-grid'') model of ISM physics, star formation, stellar
feedback, and galactic winds. In these simulations, in order to prevent
low-mass galaxies from forming too many stars, strong outflows are usually
required, which would also remove metals from the ISM.

\subsubsection{Cosmological Zoom-in Simulations: \citet{ma16mzr}}
\label{model:fire}

To improve the understanding of the physics of star formation and feedback,
\citet{ma16mzr} study the redshift evolution of the MZR using the
high-resolution cosmological zoom-in simulation FIRE \citep{hopkins14fire}. The
resolution (softening factor) of FIRE is 1--10 pc, three orders of magnitude
smaller than that of \citet{dave13} ($\sim$1 kpc). Such a high resolution
allows a realistic characterization of the physics of multi-phase ISM, star
formation, feedback, and galactic winds. FIRE includes prescriptions for a few
feedback mechanisms: (1) momentum flux from radiative pressure; (2) energy and
momentum from SN and stellar winds, and (3) photoionization and photo-electric
heating.  \citet{ma16mzr} include 22 runs of galaxies with various
star formation and merger histories.

The slope of the best-fit MZR of FIRE at $z=0.6$ (blue dashed lines in Panel
(b) of Figure \ref{fig:theo3p}, $\beta_{MZR}=0.35$) shows good agreement with
that of our MZR ($\beta_{MZR}=0.30\pm0.02$). The normalization of their MZR,
however, is lower than ours by 0.3 dex. This systematic deviation could be
physical or due to their recipe of calculating the effective metal yield,
which, as previously discussed, would lead to a 2x uncertainty. Even with this
uncertainty, their MZR matches the 1$\sigma$ confidence level of our best-fit
MZR (black dashed lines), indicating a statistical agreement.

As stated in \citet{ma16mzr}, the FIRE galaxies at \mstar$>10^{7}$\msun are
able to retain most of the metals they produced in the halos. Massive galaxies
(\mstar$>10^{10}$\msun) are even able to keep almost all of their metals. This
explains why they do not have a ``saturation'' of metallicity at the massive
end. They also find that outflows at outer radii of dark matter halos are much
less metal-enriched than those at inner radii, suggesting a high efficiency of
metal recycling.

\subsubsection{Preventative Feedback \citep{luyu15metal}}
\label{model:prevent}

In many semi-analytic models and simulations, there is a tension between
suppressing star formation and retaining enough metals in low-mass galaxies.
One way to solve it is to use preventative feedback. In contrast to ejective
scenarios where the effect of feedback is to remove gas from a galaxy to the
intergalactic medium, the preventive scenario assumes some early feedback to
change the thermal state of the intergalactic medium around dark matter halos
so that a fraction of baryons is prevented from collapsing into low-mass halos
in the first place.  Therefore, in the preventative model, the outflow strength
could be much weaker than that in ejective models. The former would result in a
higher gas-phase metallicity.

We compare the predicted MZRs of the preventative and ejective models of
\citet{luyu15metal}. The details of the models are given in
\citet{luyu15model}. We first compare their ejective model (Model-EJ in Panel
(c) of Figure \ref{fig:theo3p}) to our data. The EJ model captures most of the
common features of all ejective models in the literature. In fact, the EJ model
matches the hybrid-wind numerical simulations of \citet{dave13} very well. Both
of them match our MZR at \mstar$>10^{9.5}$\msun, but deviate from our MZR below
\mstar$=10^{9.5}$\msun\ with a steeper slope. This again demonstrates the
common issue of underpredicting metallicity for low-mass galaxies in most of
the ejective models.

The second model (Model-RI) of \citet{luyu15metal} predicts a much steeper MZR
and significantly underpredicts the {\it average} metallicity of low-mass
galaxies. This model, as an extension of the ejective model, allows the ejected
gas mass to reincorporate into the halo hot gas after a period of time.  The
reincorporated gas would decrease the gas-phase metallicity. Although the MZR
slope of this model is much steeper than our data, it is interesting to see
that the predicted metallicity is broadly consistent with those lower-branch
galaxies in our sample at $10^8$\msun$<$\mstar$<10^9$\msun\ (Panel (a) of
Figure \ref{fig:mzrsub}). This suggests that those very metal-poor galaxies may
have experienced significant re-infallings of their ejected gases. Future work,
including robust measurement of the metal-poor galaxies and more accurate
modeling, is required to investigate the formation of very metal-poor galaxies.

On the other hand, the preventative model (Model-PR in the same panel)
overpredicts the metallicity of low-mass galaxies. Because the main mechanism
responsible for keeping baryon mass low in low-mass galaxies is to prevent
baryons from collapsing into their host halos, outflow in this model is
moderate.  Therefore, galaxies in this model retain a larger fraction of
metals. The model predicts a rather flat MZR at $z=0.6$, with the metallicity
of low-mass (\mstar$\sim10^{8}$\msun) galaxies higher than our data by 0.6 dex.

This result suggests that a pure preventative model cannot explain the MZR at
$z=0.6$.  Comparing the model predictions with our observational result, we
find that the observed MZR sits between the predictions of the Model-EJ and the
Model-PR, suggesting that both ejection and prevention work together in
low-mass galaxies. Other observations also show evidence of strong outflows
\citep[e.g.,][]{rubin14}, indicating that ejection works at some level, but the
effect of outflow in removing hydrogen mass and metal mass is yet to be
measured quantitatively. \citet{luyu15metal} also shows that the pure
preventative feedback model is able to match the MZR of $z=0$ SDSS galaxies,
but overestimates metallicity for $z=2$ galaxies. Therefore, the importance of
preventative feedback may also evolve with redshift.

Last but not least, we point out that the predictions made in
\citet{luyu15metal} should be considered as upper limits for each scenario.
The authors assumed that the metallicity of outflow is the same as the
gas-phase metallicity of a galaxy. If the outflow material is more metal
enriched relative to the ISM because outflow is driven by SNe, which are the
sources of metals, the predicted MZR would decrease. 

\subsection{Summary of Model Comparisons Using Slope and Normalization}
\label{model:summary}

In this section, we compare our MZR at $0.5 \leq z \leq 0.7$ to a variety of
theoretical works, from simple scaling relation to state-of-the-art numerical
simulations.  Here we summarize the comparisons of using the slope and
normalization of the MZR.

\begin{enumerate} 

\item {\it Slope}. Models \citep[][and D12]{dekel03,forbes14mzr} incorporating
SN energy-driven wind (with mass loading factor $\eta \propto
M_{halo}^{(-2/3)}$) provide good agreement with the slope of the MZR of
low-mass (\mstar$<10^{9.5}$\msun) galaxies. For massive galaxies, gas recycling
(the $\alpha_Z$ term or its analogs) plays an important role, but the
characterization of $\alpha_Z$ is uncertain for model comparison. The latest
high-resolution simulation FIRE \citep{hopkins14fire,ma16mzr}, which has the
ISM physics more accurately characterized thanks to its high resolution,
produces a slope in good agreement with ours across the whole \mstar\ range in
our paper.

\item {\it Normalization}. With the uncertainty of metal yield $y$ in mind, we
find that the hybrid-wind simulation of \citet{dave13} and the ejective model
of \citet{luyu15metal} match the normalization of our MZR at the massive end.
These models, together with \citet{mitra15}, underestimate the metallicity of
low-mass galaxies. One possible solution is to mix preventative
\citep{luyu15model,luyu15metal} and ejective feedback for low-mass galaxies.

\item {\it Uncertainties}. Our model comparisons are subjected to uncertainties
from both observational and theoretical sides. The largest uncertainty of
observations is the calibration uncertainty, namely the uncertainty of mapping
emission line ratios to metallicities. Two major uncertainties of theoretical
models are the metal yield, which can be strongly modulated by the metal
enrichment of the outflow material relative to the ISM, and the gas recycling
term for massive galaxies. 

\end{enumerate}

\subsection{Using Slope of the MZR to Link to Dark Matter Halos}
\label{model:halo}

The observed slope of the MZR could also reveal new insights in the connection
between the luminous (baryonic) and dark (dark matter halo) sides of galaxy
formation. This connection can be established through simple analytic models,
which usually contain much less parameters than semi-analytic models and even
hydrodynamical simulations. Some of these parameters are hard to constrain, as
discussed in previous sub-sections. Here, we use the model in \citet{lilly13}
to demonstrate the efficiency of such analytic models.

The model of \citet{lilly13} connects three different aspects of galaxy
formation and evolution: (1) evolution of SSFR relative to the growth of dark
matter halos, (2) gas-phase metallicities of galaxies, and (3) \mstar--\mhalo\
relation. In the model, the formation of stars is instantaneously regulated by
the mass of gas in a varying reservoir. The gas in the reservoir is controlled
by gas inflow into galaxy and outflow expelled from galaxy, the latter of which
is in turn scaled with the SFR. 

In the model, gas-phase metallicity is linked to dark matter halos through
$f_{gas}$, the ratio of the reduced SFR (i.e., SFR with only long-lived stars
counted) to the gas infall rate. Assuming the metallicity of inflow is
negligible, we have $Z = f_{gas}y$,
where $y$ is the yield. $f_{gas}$ is proportional to the \mstar--\mhalo\
relation: $f_{gas} \propto {\rm M_*}/{\rm M_{halo}}$.  Therefore, the slope of
the MZR ($Z \propto {\rm M_*}^{\beta_{MZR}}$) and the \mhalo--\mstar\ relation
(${\rm M_{halo}} \propto {\rm M_*}^\Upsilon$) has a relation (which is a
rearrangement of Equations (32) and (34) of \citet{lilly13}):

\begin{equation}
\label{eq:zhalo}
\Upsilon = 1-\beta_{MZR},
\end{equation}
where $\Upsilon$ has the same definition as that used in Equation
\ref{eq:beta_low} and \ref{eq:beta_high}. 

In Section \ref{model:winds}, we simply take $\Upsilon \sim 0.5$ from
\citet{moster10} to predict model MZR slopes. Here, we explore the link between
the MZR and dark matter halos in the other way: starting from our observed
$\beta_{MZR}$ and using it to constrain the \mstar--\mhalo\ relation. Using
Equation \ref{eq:zhalo}, our observed $\beta_{MZR}$=0.30 yields
$\Upsilon$=0.70. This value is larger than that of \citet{moster10} and hence
implies a {\it flatter} (\mstar/\mhalo)--\mhalo\ relation than that in
\citet{moster10}. It is, however, interesting to note that at
\mhalo$<10^{11}$\msun, the slope of the (\mstar/\mhalo)--\mhalo\ relation in
\citet{behroozi13} is indeed flatter than that in \citet{moster10} (see Figure
14 in \citet{behroozi13}) and quite similar to what inferred from our
$\Upsilon$=0.7. Our example here simply aims to highlight the power of using
$\beta_{MZR}$ to constrain the physics of dark matter halos. Future
investigations are needed to better determine the slope of the
(\mstar/\mhalo)--\mhalo\ relation at the very low-mass end.

\subsection{Using Scatter of the MZR to Understand the Formation of Low-mass Galaxies}
\label{model:scatter}

The scatter of the MZR is crucial to understand the origin of the MZR. Relative
to the slope and normalization, the scatter of the MZR is barely affected by
the calibration uncertainty \citep[e.g.,][]{zahid12}. Therefore, theoretical
work on modeling the scatter may provide important clues to understanding
galaxy formation. Here, we use the statistical equilibrium model of
\citet{forbes14mzr} to demonstrate how to use the scatter to explore the
stochastic nature of the formation of low-mass galaxies.

Although the equilibrium models with energy-driven winds provide a good
explanation of the slope of the MZR at the low-mass end, its explicit
assumption $dM_{gas}/dt = 0$ may not be true for all galaxies. For example,
\citet{feldmann13} and \citet{forbes14disk} point out that many galaxies are
out of the equilibrium with $dM_{gas}/dt < 0$. On the other hand, in
semi-analytic models \citep[e.g.,][]{luyu14candels}, low-mass galaxies always
increase their gas mass rapidly, namely, $dM_{gas}/dt > 0$. Moreover, assuming
$dM_{gas}/dt = 0$ for all galaxies only allows one to derive the first-order
scaling relations; it cannot shed light on the origin of the scatter of the
scaling relations.

\citet{forbes14mzr} present a simple model to understand the origin of the
scatter in star formation and metallicity of galaxies at a fixed mass. This
model relaxes the key assumption of equilibrium models, namely that the rate at
which baryons enter the gas reservoir varies slowly. Galaxies in this model
have been fed by some stochastic accretion process long enough that the full
joint distribution of all galaxy properties has become time invariant. This
model can be referred to as a statistical equilibrium model because individual
galaxies are not in equilibrium, but the population is. The scatter of scaling
relations arises from the intrinsic scatter in the accretion rate and also
depends on the time-scale on which the accretion varies compared to the
time-scale on which the galaxy loses gas mass. 

\citet{forbes14mzr} use $\gamma = 2/3$ (see our Equation \ref{eq:eta}) for the
mass loading factor, consistent with energy-driven winds, and they do not
include any wind recycling.  Given our discussions in Section
\ref{model:winds}, not surprisingly, their MZR slope ($\beta_{MZR}=0.348$, the
light green line in Panel (a) of Figure \ref{fig:theo3p}) matches that of the
energy-driven wind ($\beta_{MZR}=0.33$, the magenta line).  Their slope agrees
with our slope ($0.30\pm0.02$) for all \mstar\ ranges within 3$\sigma$ level.
These results again suggest the importance of SN feedback on shaping the MZR.

More interesting is to compare the scatter of the MZR in the models of
\citet{forbes14mzr} to our data (Panel (b) of Figure \ref{fig:scatter}).
\citet{forbes14mzr} do not tune their models to obtain a set of ``best-fit''
parameters. Instead, they only use two sets of parameters: initial guess and
improved guess. The difference between the two sets of parameters allows us to
explore the processes responsible for the origin of the scatter of the MZR. The
key parameters that govern the scatter of the MZR in their model are (1) the
scatter of baryonic accretion rate ($\sigma_{\dot{M}}$) and (2) the scatter of
\mstar--\mhalo\ relation ($\sigma_{SHMR}$).

The parameters in the initial guess are taken from N-body simulations of
\citet{neistein08} and \citet{neistein10}. The initial guess yields an MZR
scatter of 0.16 (solid gray line in Panel (b) of Figure \ref{fig:scatter}),
which matches our scatter around \mstar$=10^{9}$\msun. To produce a scatter
smaller than the observed scatter of the local MZR to approach the intrinsic
scatter, \citet{forbes14mzr} adjust their key parameters by reducing
$\sigma_{\dot{M}}$ and $\sigma_{SHMR}$ by half. The improved guess yields an
MZR scatter of 0.09 (dotted gray line in Panel (b) of Figure
\ref{fig:scatter}), now matching our MZR scatter of massive galaxies.

The adjustment of the parameters in \citet{forbes14mzr} has an important
implication on the origin of the MZR scatter: to increase the MZR scatter, one
can increase $\sigma_{\dot{M}}$ and/or $\sigma_{SHMR}$. Galaxies out of
(statistical) equilibrium ($dM_{gas}/dt > 0$) could also increase the MZR
scatter, as argued by \citet{dave11a} and \citet{zahid12} that the MZR scatter
would be large if the timescales for galaxies to equilibrate is long (namely,
gas dilution time scale is long compared to the dynamical time scale of
galaxies).

\citet{forbes14mzr} predict a constant MZR scatter over the whole \mstar\ range
in both the initial and improved guesses, but our observed scatter increases as
\mstar\ decreases. The difference stems from the assumption in
\citet{forbes14mzr} that both $\sigma_{\dot{M}}$ and $\sigma_{SHMR}$ are mass
independent. In fact, many current abundance matching methods use a constant
$\sigma_{SHMR}$ in their models \citep[e.g.,][]{moster10,behroozi13}. There is
some evidence of a mass-independent $\sigma_{SHMR}$ for massive galaxies with
\mstar$>10^{10.2}$\msun\ \citep{trujillogomez11,reddick13}, but $\sigma_{SHMR}$
of low-mass galaxies has not been well constrained. 

The hybrid-wind simulations of \citet{dave13} also predict an increasing
scatter toward the low-mass end at $z=0.55$ (the solid gray line in Panel (b)
of Figure \ref{fig:scatter}). Their scatter matches ours very well for
\mstar$>10^{9}$\msun\ galaxies. Below \mstar$=10^{9}$\msun, their scatter
becomes smaller than ours. In \citet{dave13}, since the baryonic accretion rate
is well set by N-body simulation and their galaxies are probably already out of
equilibrium (see Section \ref{model:ezw}), one possible explanation of their
smaller-than-observed scatter below \mstar$=10^{9}$\msun\ is that their
$\sigma_{SHMR}$ is smaller than it should be. An additional piece of evidence
of an underestimated $\sigma_{SHMR}$ of \citet{dave13} is their
tighter-than-observed SFR--\mstar\ relation. In the data of \citet{dave13},
there are almost no galaxies with SFR$>1 M_\odot/yr$ at \mstar$<10^{9}$\msun.
In contrast, our sample contains such galaxies (see Figure \ref{fig:sample}).

The lack of redshift evolution of the scatter of the MZR is also intriguing.
As shown by Panel (b) of Figure \ref{fig:scatter}, our scatter at $z\sim0.7$ is
consistent with that of \citet{tremonti04} at $z\sim0$. This result of no
redshift evolution implies that all the factors that could alter the scatter of
the MZR should remain unchanged from $z\sim0.7$ to $z\sim0$. Among them,
$\sigma_{\dot{M}}$ is particularly interesting. Our results suggest that,
although the cosmic gas accretion rate decreases by a factor of three during
this period \citep[e.g.,][]{dekel09gas}, the scatter of the accretion rate
remains unchanged from $z\sim0.7$ to $z\sim0$. Future theoretical studies on
the redshift evolution of the scatter of baryonic accretion rate is important
to understand the origin of the scatter of the MZR.

\section{Conclusions}
\label{conclusion}

We study the MZR and its scatter at $0.5\leq z\leq0.7$ by using 1381 field
galaxies collected from previous deep spectroscopic surveys. Our sample is
fairly representative of normal star-forming galaxies at the redshift, in terms
of SSFR and color at a given \mstar. Moreover, the sample contains 237
galaxies with \mstar$<10^{9}$\msun, comprising currently the largest sample in
this mass regime ($\sim$10 times larger than previous ones) beyond the the
local universe, which enables an unprecedentedly strong constraint on the MZR
and its scatter in the low-mass regime.

We find a power-law MZR with a slope (in logarithmic space) of 0.30$\pm$0.02 at
$10^{8}$\msun$<$\mstar$<10^{11}$\msun. Our MZR shows agreement with other
MZRs at similar redshifts in the literature at \mstar$>10^{9}$\msun. The slope
of our MZR below \mstar$\sim10^{9}$\msun\ is flatter than the extrapolation of
other MZRs. The SFR dependence of the MZR in our sample is weaker than that in
the local universe. More tests are needed to investigate the existence of the
fundamental metallicity relation beyond the local universe.

We compare our MZR to several theoretical models, including simple scaling
relations, semi-analytic models, and state-of-the-art numerical
simulations. We find that models incorporating SN energy-driven winds (with
mass loading factor $\eta \propto M_{halo}^{(-2/3)}$) provide good agreement
with the slope of the MZR of galaxies with \mstar$<10^{9.5}$\msun.

With the 10-fold gain in sample size, we present the first measurement of the
scatter of the MZR down to \mstar$=10^{8}$\msun\ at $0.5 \leq z \leq 0.7$. The
scatter increases as \mstar\ decreases, from 0.1 dex at
\mstar$\sim10^{10}$\msun\ to 0.3 dex at \mstar$\sim10^{8}$\msun. The scatter of
the MZR shows no evolution from $z\sim0.7$ to $z\sim0$. 

Relative to the slope and normalization of the MZR, which are subjected to both
observational and theoretical uncertainties, the scatter of the MZR is the
least affected by observational uncertainties and thus can be used as an
important diagnostic of the stochastic formation history of low-mass galaxies.
According to a simple statistical equilibrium model, the large scatter in
low-mass galaxies implies that either $\sigma_{\dot{M}}$ or $\sigma_{SHMR}$
increases as \mstar\ decreases. The lack of the redshift evolution of the
scatter implies that both $\sigma_{\dot{M}}$ and $\sigma_{SHMR}$ remain
unchanged from $z=0.7$ to $z=0$.

\ \ \

We thank the anonymous referee for constructive comments that improve this
article. We thank Aldo Rodriguez-Puebla for useful discussions. Several authors
from UCSC acknowledge support from NSF grant AST-0808133. Support for Program
HST-GO-12060 and HST-AR-13891 were provided by NASA through a grant from the
Space Telescope Science Institute, which is operated by the Association of
Universities for Research in Astronomy, Incorporated, under NASA contract NAS
5-26555. MR also acknowledges support from an appointment to the NASA
Postdoctoral Program at Goddard Space Flight Center. JF is supported by
HST-AR-13909. JRT acknowledges support from NASA through Hubble Fellowship
grant \#51330 awarded by the Space Telescope Science Institute. AD is supported
by ISF grant 24/12, by the I-CORE Program of the PBC ISF grant 1829/12, and by
NSF grant AST-1405962. PGPG acknowledges support from Spanish MINECO grant
AYA2012-31277.

{\it Facilities}: Keck\ (DEIMOS)

\appendix

\twocolumngrid

\section{A. Possible Selection and Measurement Effects on Our MZR}
\label{appendix:effect}

\begin{figure*}[htbp] 
\includegraphics[scale=0.20,angle=0,clip]{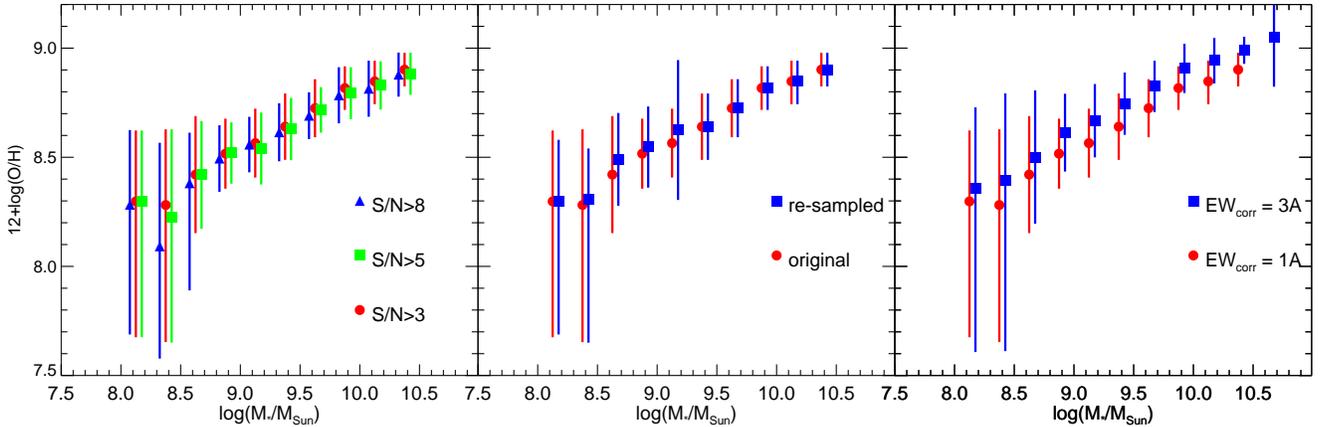}

\caption[]{Selection and measurement effects on our MZR measurement. In all
panels, red circles with error bars are the median and 16th and 84th
percentiles of the DEEP3+TKRS ``upper+lower Z'' MZR with ${\rm S/N_{[OIII] and
H\beta}}>$3 and H$\beta$ EW correction of 1 \AA\ (the same as that in Panel (a)
of Figure \ref{fig:mzrsub}). {\bf Left}: S/N cut effect. Blue triangles and
green squares show the re-calculated MZRs with ${\rm S/N_{[OIII] and
H\beta}}>$8 and 5, respectively. {\bf Middle}: Sample bias effect. Blue squares
and error bars show the MZR of re-sampling our DEEP3+TKRS sample to match the
SSFR distribution at a given \mstar\ of our parent samples. This panel tests
the effect of our final sample being biased toward slightly high-SSFR galaxies
(see Figure \ref{fig:sample}). {\bf Right}: H$\beta$ EW correction effect. Blue
squares and error bars show the MZR calculated with H$\beta$ EW correction
factor of 3 \AA.

\label{fig:effect}} 
\end{figure*}

Here we discuss three selection and measurement effects that may have impacts
on our MZR measurement: (1) S/N cuts on [OIII] and H$\beta$, (2) our sample
bias toward higher-SSFR galaxies at \mstar$<10^{9.5}$\msun, and (3) our choice
of an H$\beta$ absorption correction factor of EW = 1 \AA. We use both the
DEEP3+TKRS ``upper+lower Z'' MZR and the [OIII]/[OII]-derived MZR as our
fiducial MZR for the tests. The test results do not change with the choice of
the fiducial MZR. For simplicity, we only show the results of the ``upper+lower
Z'' MZR in Figure \ref{fig:effect}. We conclude that none of the above effects
would significantly change the slope and scatter of our MZR.  

To investigate the effect of the S/N cuts of emission lines, we re-calculate
the MZR with higher S/N cuts, namely ${\rm S/N_{[OIII], H\beta}}>$5 and 8.  The
new MZRs in the left panel of Figure \ref{fig:effect} show that neither cut
significantly changes our fiducial results with S/N$>$3. On average, the
deviation between the new (blue and green) and the fiducial (red) MZRs is about
0.05 dex. The only large deviation is found at
$10^{8}$\msun$<$\mstar$<10^{8.5}$\msun\ with S/N$>$8, but the error bars in
this mass regime are also the largest. This result is consistent with
\citet{foster12}, who also found that varying the selection criteria of S/N cut
or magnitude cut does not significantly alter the MZR for a given calibration.
Therefore, we conclude that the S/N cut only induces minor effects on our MZR. 

We also investigate the effect of our sample bias on the MZR. As shown in
Figure \ref{fig:sample}, our sample is biased toward high-SSFR galaxies at
\mstar$<10^{9.5}$\msun. To derive the MZR for a mass-complete sample, we assign
an [OIII]/H$\beta$ value to each star-forming galaxy in our parent samples
(CANDELS GOODS-N and IRAC EGS, shown by contours and small black dots in Figure
\ref{fig:sample}). The assigned value is equal to the line ratio of its closest
galaxy in the (\mstar, SSFR) space in our final sample (red symbols in Figure
\ref{fig:sample}). The assumption here is that the [OIII]/H$\beta$ value is
determined by its \mstar\ and SSFR. This assignment is feasible because our
selected sample, although biased, actually covers the whole star-forming main
sequence. We then re-calculate the MZR for the whole parent sample with the
assigned line ratios.

The result (middle panel of Figure \ref{fig:effect}) shows that the bias in
SSFR induces almost no effect on the MZR. This is consistent with our results
that the MZR dependence on SSFR is weak in our sample in Section
\ref{metal:sfr}, where the most obvious (but still weak) signal of SSFR
dependence of metallicity is found at $10^{8.5}$\msun$<$\mstar$<10^{9.5}$\msun\
(see Figure \ref{fig:fmr}). Therefore, we conclude that our sample is
representative of star-forming galaxies for deriving the MZR and its scatter
between $10^{8}$\msun\ and $10^{11}$\msun. 

We also test the effect of EW correction for H$\beta$ absorption. We assume a
correction factor of 1 \AA\ for our galaxies. Some authors, however, found a
higher correction factor (for example, 3 \AA\ in \citet{lilly03}. We
re-calculate the MZR using a correction factor of EW=3\AA. The result (right
panel of Figure \ref{fig:effect}) shows that the metallicity of galaxies with
$10^{8.5}$\msun$<$\mstar$<10^{10.5}$\msun\ is increased by $\sim$0.1 dex.
What's important is that although the increase changes the normalization of the
MZR, it does not significantly change its slope or scatter. We emphasize that
we use the correction of EW=1 \AA\ as our fiducial results because it is drawn
from previous observations with similar spectral resolution as ours. As
discussed in \citet{zahid11}, the EW correction factor depends on the spectral
resolution. Lower resolution requires larger correction factors because
emission lines are spread into larger wavelength regions. Also, comparison with
previous MZRs (Panel (d) of Figure \ref{fig:mzrsub}) also suggests that using
EW=1 \AA\ yields a better agreement with previous results in the literature.

\section{B. Effect of AGN Removal}
\label{appendix:agn}

\begin{figure}[htbp]
\includegraphics[scale=0.45,angle=0,clip]{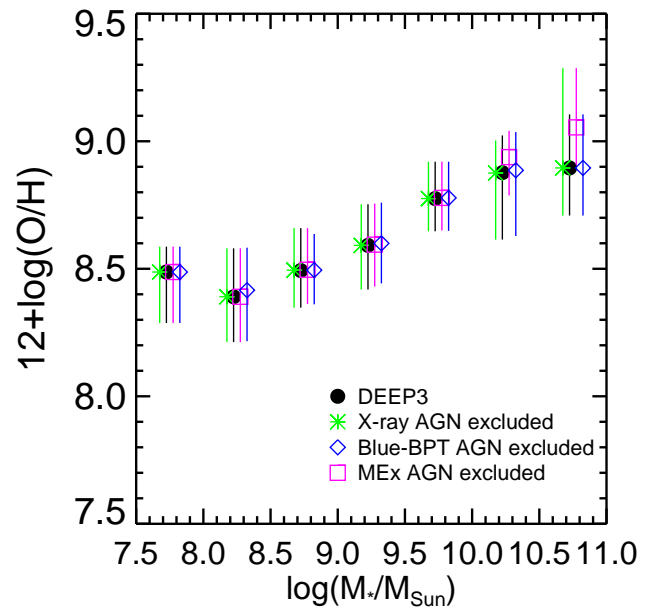}

\caption[]{Effect of AGN removal on the MZR measurement. For simplicity, all
galaxies are assumed to be in the upper branch, and no conversion uncertainty
is included. Large black circles with error bars show DEEP3+TKRS galaxies
without any AGN removal. MZRs with different AGN removal methods are shown by
different symbols as indicated by the labels. The MEx method (purple squares)
is the fiducial method used in our paper.

\label{fig:agn}}
\end{figure}

As discussed in Section \ref{metal:compare}, AGN removal affects the slope of
MZR at the massive end. In this paper, we use the MEx method of
\citet{juneau11} to exclude AGN contamination. \citet{juneau11} used the DEEP2
and TKRS spectra (the same data as used by this paper) to calibrate their
method to achieve a balance between efficiency and contamination. Also, the MEx
method is shown remaining effective up to $z=1.5$ \citep{trump13}. Therefore,
we believe that the MEx method is the most suitable one for our study. In
Figure \ref{fig:agn}, we re-calculate the MZR by using other AGN removal
methods and compare the results to our fiducial MZR of using the MEx method.

X-ray is the most reliable way to identify AGNs, but it is not complete, due to
a significant (up to ~50\%) fraction of Compton-thick AGNs. Therefore, the
results of using X-ray selection should only be treated as a lower limit of AGN
contamination. As shown in Figure \ref{fig:agn}, X-ray selection removes fewer
high-metallicity (i.e., high [OIII]/H$\beta$) galaxies at \mstar$>10^{10}$\msun
than the MEx method does, resulting in an almost flat MZR at this mass regime.
The massive-end slope is now more consistent with that of \citet{zahid13} 
because \citet{zahid13} also only removed X-ray sources. 

AGN removal can also be done with the ``Blue BPT'' diagram
\citep{lamareille04,lamareille10} which uses [OIII]/H$\beta$ vs. [OII]/H$\beta$
to identify AGNs. Compared to the MEx method, the ``Blue BPT'' introduces no
explicit mass dependence. It, however, requires measurement of dust extinction
because [OII]/H$\beta$ is reddening dependent. \citet{lamareille10} argues that
using EW ratios could alleviate the issue, but this does not fully remove the
reddening-dependence because stars and gas have different extinction. Figure
\ref{fig:agn} shows that the ``Blue BPT'' removal results in a very similar MZR
to the MEx (and X-ray) removal at \mstar$<10^{10.5}$\msun. At higher \mstar
(where small number statistics hits our sample), the ``Blue BPT'' result is
similar to that of X-ray removal.

Overall, we conclude: (1) X-ray selection provides the most reliable AGN
identification, but its identified AGN sample is not complete. Line-ratio
diagnostics are needed to exclude Compton-thick AGNs. (2) MEx of
\citet{juneau11} is defined at $z\sim0.7$ and hence the most suitable method
for our study. (3) ``Blue BPT'' yield a similar result to X-ray selection. (4)
What's important is that the three methods result in almost same MZR at
\mstar$<10^{10.5}$\msun. This is because AGN contamination is very little in
low-mass galaxies \citep{trump15}. Therefore, our main conclusions on both the
slope and scatter of MZR at \mstar$<10^{10.5}$\msun\ are unaffected by our AGN
removal.

\bibliographystyle{apj}

\end{document}